\documentclass{pasa}

\usepackage{graphicx}
\usepackage{xspace}

\title[The triple nucleus of Sp~3]{Two's company, three's a crowd: SALT reveals the likely triple nature of the nucleus of the extreme abundance discrepancy factor planetary nebula Sp 3\thanks{Based on observations made with the Southern African Large Telescope (SALT) under programmes 2012-1-RSA\_OTH\_010, 2016-2-SCI-034 and 2017-1-MLT-010.}}

\author[Miszalski et al.]{Brent Miszalski$^{1,2,6}$, Rajeev Manick$^{1}$, Thomas Rauch$^{3}$, Krystian I{\l}kiewicz$^{4}$, Hans Van Winckel$^{5}$ and Joanna Miko{\l}ajewska$^{4}$ 
\affil{$^1$South African Astronomical Observatory, PO Box 9, Observatory, 7935, South Africa}
\affil{$^2$Southern African Large Telescope Foundation, PO Box 9, Observatory, 7935, South Africa}
\affil{$^3$Institute for Astronomy and Astrophysics, Kepler Center for Astro and Particle Physics, Eberhard Karls University, Sand 1, 72076 T\"ubingen, Germany}
\affil{$^4$Nicolaus Copernicus Astronomical Center, Polish Academy of Sciences, Bartycka 18, PL-00716 Warsaw, Poland}
\affil{$^5$Instituut voor Sterrenkunde, KU Leuven, Celestijnenlaan 200D bus 2401, B-3001 Leuven, Belgium}
\affil{$^6$Email: brent@saao.ac.za}
}

\jid{PASA}
\doi{10.1017/pas.\the\year.xxx}
\jyear{\the\year}

\usepackage{aas_macros}

\newcommand{\arcsec}{''}
\newcounter{Rco}

\newcommand{\logg}{\mbox{$\log g$}\xspace}
\newcommand{\loggw}[1]{\mbox{$\log g\hspace{-0.5mm} =\hspace{-0.5mm}  #1$}}

\newcommand{\sga}{\raisebox{-0.10em}{$\stackrel{>}{{\mbox{\tiny $\sim$}}}$}}

\newcommand{\sla}{\raisebox{-0.10em}{$\stackrel{<}{{\mbox{\tiny $\sim$}}}$}}

\newcommand{\Teff}{\mbox{$T_\mathrm{eff}$}\xspace}
\newcommand{\Teffw}[1]{\mbox{$\Teff\hspace{-0.5mm} =\hspace{-0.5mm} #1 \,\mathrm{K}$}}
\newcommand{\ebv}{\mbox{$E_\mathrm{B-V}$}}

\newcommand{\Msol}{$M_\odot$\xspace}

\newcommand{\mmspr}{\hbox{}\hspace{+0.8cm}}
\newcommand{\smspr}{\hbox{}\hspace{+2.5mm}}

\begin{document}

\begin{frontmatter}
\maketitle

\begin{abstract}
   The substantial number of binary central stars of planetary nebulae (CSPNe) now known ($\sim$50) has revealed a strong connection between binarity and some morphological features including jets and low-ionisation structures. However, some morphological features and asymmetries might be too complex or subtle to ascribe to binary interactions alone. In these cases a tertiary component, i.e. a triple nucleus, could be the missing ingredient required to produce these features. The only proven triple, NGC~246, is alone insufficient to investigate the shaping role of triple nuclei, but one straight-forward way to identify more triples is to search for binaries in nuclei with known visual companions. Here we demonstrate this approach with the SALT HRS discovery of a 4.81 d orbital period in the CSPN of Sp~3 which has a visual companion 0.31\arcsec\, away. The spectroscopic distance of the visual companion is in agreement with distance estimates to the nebula, the \emph{GAIA} DR2 parallax of the central star, and the gravity distance of the central star. This supports a physical association between the visual companion and the inner 4.81 d binary, making the nucleus of Sp~3 a likely triple. We determine $T_\mathrm{eff}=68^{+12}_{-6}$ kK, $\log g=4.6\pm0.2$ cm s$^{-2}$ and $v_\mathrm{rot}=80\pm20$ km s$^{-1}$ for the primary from NLTE model atmosphere analysis. The peculiar nebula presents an apparent bipolar morphology, jets and an unexpected `extreme' oxygen abundance discrepancy factor (adf) of 24.6$^{+4.1}_{-3.4}$. The adf is inconsistent with the purported trend for longer orbital period post-CE PNe to exhibit normal adfs, further highlighting the dominant influence of selection effects in post-CE PNe. Lastly, the Type-I nebular abundances of Sp~3, whose origin is often attributed to more massive progenitors, are incongruous with the likely Galactic Thick Disk membership of Sp~3, possibly suggesting that rotation and binarity may play an important role in influencing the AGB nucleosynthesis of PNe. 
\end{abstract}

\end{frontmatter}

\section{Introduction}
\label{sec:intro}
Binary interactions are fundamental to understand the formation of planetary nebulae (PNe) and their diverse characteristics (De Marco 2009; Jones \& Boffin 2017a). Observational studies are beginning to probe how binary central stars of PNe (CSPNe) influence the shape of their surrounding nebulae. The most commonly observed binaries in PNe are main-sequence or white dwarf (WD) stars orbiting the WD primary in $\sim$1 d or less. These binaries have recently emerged from a common-envelope (CE) phase (Ivanova et al. 2013) and occur in around 1 in 5 PNe (Bond 2000; Miszalski et al. 2009a). Observations of post-CE PNe have shown that aspects of the nebula morphology were directly influenced by the binary interaction that created the PN.  These aspects include the creation of accretion driven precessing outflows or jets (Boffin et al. 2012; Miszalski et al. 2013; Tocknell et al. 2014 and ref. therein) and alignment of the nebula orientation with orbital inclination (Hillwig et al. 2016). Low-ionisation filaments also appear to be associated with post-CE PNe (Miszalski et al. 2009b; Miszalski et al. 2011a, 2019a), particularly in ring configurations (e.g. Corradi et al. 2011; Boffin et al. 2012; Miszalski et al. 2018a). Miszalski et al. (2009b) suggested these rings were the result of a photoionising wind interacting with material deposited during the CE phase and this interpretation was recently supported by simulations (Garc\'ia-Segura et al. 2018). Other characteristics of binarity may also be a tendency for large abundance discrepancy factors (Wesson et al. 2018 and ref. therein) and bipolar geometries (Miszalski et al. 2009b; Miszalski et al. 2018b), however the precise conditions responsible for producing these characteristics remain unclear. 

The extent to which companions at larger orbital separations on the order of $\sim$1-1000 au could shape the surrounding nebula is more uncertain. Several studies have focused on how these systems may shape nebulae (e.g. Soker 1994, 1999; Soker \& Rappaport 2000; Gawryszczak et al. 2002; Kim \& Taam 2012), but there is a paucity of observed systems to compare against these predictions. The pioneering work of Ciardullo et al. (1999) used the \emph{Hubble Space Telescope} to discover 10 probable, 6 possible and 3 doubtful visual companions to CSPNe with very large separations in excess of 100 au. Proving a physical association for these candidates requires additional observations. Apart from the Ciardullo et al. (1999) sample, there are few other PNe with promising visual companions (Bobrowsky et al. 1998; Benetti et al. 2003; Liebert et al. 2013; Adam \& Mugrauer 2014). More recently, four binaries with orbital separations intermediate between post-CE and visual binaries were discovered via radial velocity (RV) monitoring (Van Winckel et al. 2014; Jones et al. 2017; Miszalski et al. 2018a). Further studies of these large orbital separation binaries and surveys for new examples are necessary to better understand their potential role in shaping PNe. 

A corollary of efforts to identify larger orbital sepration companions in PNe is that triple or higher order multiple systems (Toonen et al. 2016) are much more accessible for discovery. Triple systems are expected to occur in PNe if they derive from main sequence triples (De Marco 2009) and could potentially explain the more complex or so-called ``messy'' PNe morphologies (e.g. Soker et al. 1992; Soker 2016; Bear \& Soker 2017). The only confirmed triple belongs to NGC~246 in which the PG1159 type primary has two comoving companions, each with spectral types of M5-6V and K2-5V with projected separations from the primary of $\sim$500 au and $\sim$1900 au, respectively (Adam \& Mugrauer 2014). Another similar triple may also be present in NGC~7008 and requires confirmation (Ciardullo et al. 1999). In two other cases further observations might be able to reclassify a known binary as a triple. Ciardullo et al. (1999) found a $V=15.87$ mag star separated 2.82\arcsec\ from the binary nucleus of A~63 ($P=0.46$ d, Bond et al. 1978), although distance estimates suggest it is more likely a foreground star (Ciardullo et al. 1999). Jones et al. (2017) suggested another star may be necessary to explain the unexpectedly high primary mass in the binary nucleus of LoTr~5 ($P=2717\pm63$ d, Van Winckel et al. 2014; Jones et al. 2017). Other proposed triples include M~2-29 (Hajduk et al. 2008) and SuWt~2 (Exter et al. 2010 and ref. therein), but neither withstand further scrutiny (Miszalski et al. 2011b; Jones \& Boffin 2017b).

Adam \& Mugrauer (2014) utilised high resolution imaging to prove the triple nature of NGC~246. An alternative approach is to identify the presence of a third star in known spectroscopic or visual binaries. Following this approach we present an observational study of the PN Sp~3 (PN G342.5$-$14.3) which was included in an ongoing, systematic survey to search for long-period binary central stars of PNe (Miszalski et al. 2018a, 2018b, 2019b) with the Southern African Large Telescope (SALT, Buckley et al. 2006; O'Donoghue et al. 2006). Sp~3 is a relatively unstudied PN notable for the probable association between the $V=13.20$ mag central star and a $V=16.86$ mag visual companion (Ciardullo et al. 1999). As in the case of NGC~1360 (Miszalski et al. 2018a) and NGC~2392 (Miszalski et al. 2019a), Af{\v s}ar \& Bond (2005) detected radial velocity (RV) variability in nine observations of Sp~3, but did not determine an orbital period. Section \ref{sec:obs} describes the imaging and spectroscopic observations taken with SALT which are analysed in Section \ref{sec:analysis}. We discuss the results in Section \ref{sec:discussion} and conclude in Sect. \ref{sec:conclusion}. 

\section{OBSERVATIONS}
\label{sec:obs}
\subsection{Narrow-band imaging}
\begin{figure*}
   \begin{center}
      \includegraphics[scale=0.45,bb=0 0 512 512]{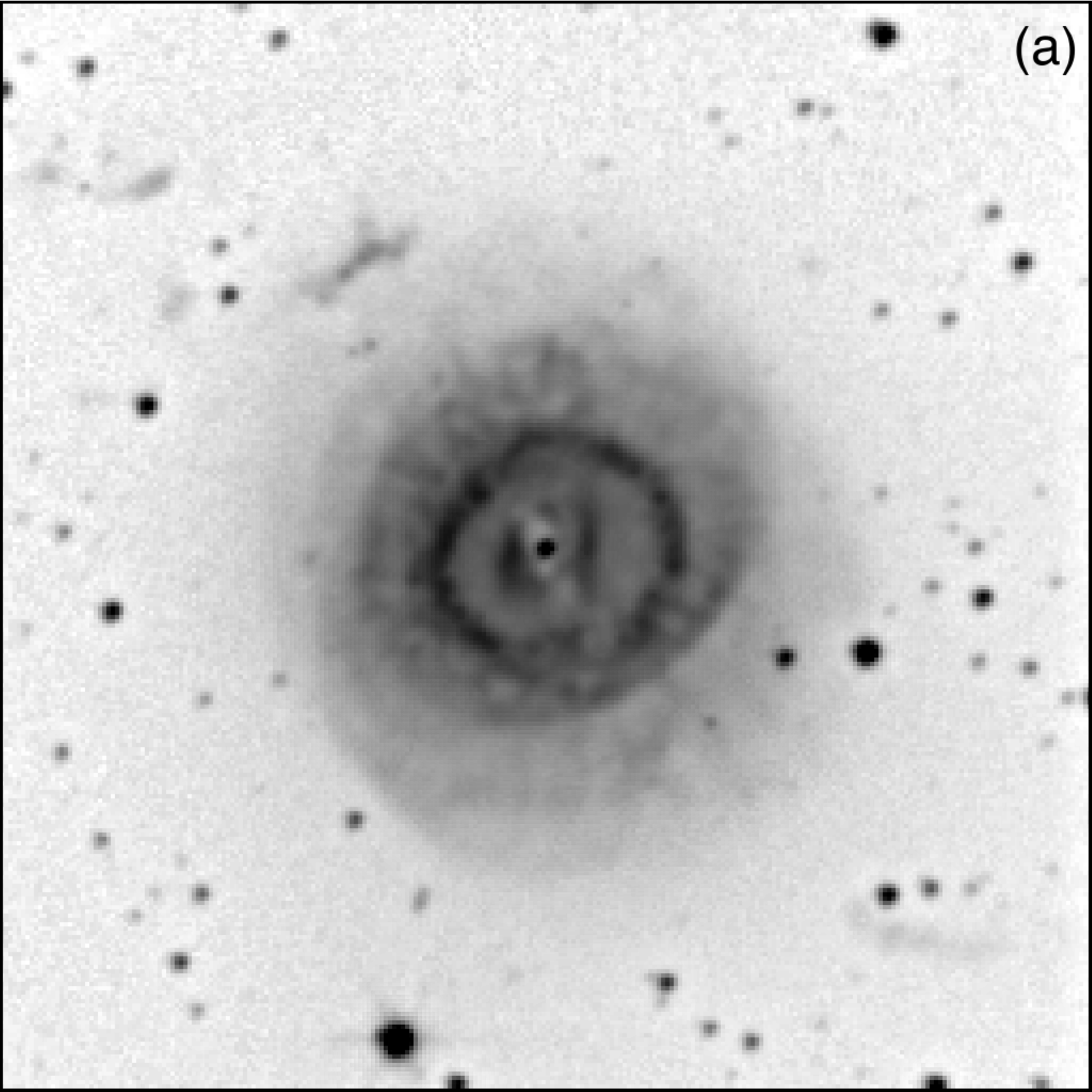}
      \includegraphics[scale=0.45,bb=0 0 512 512]{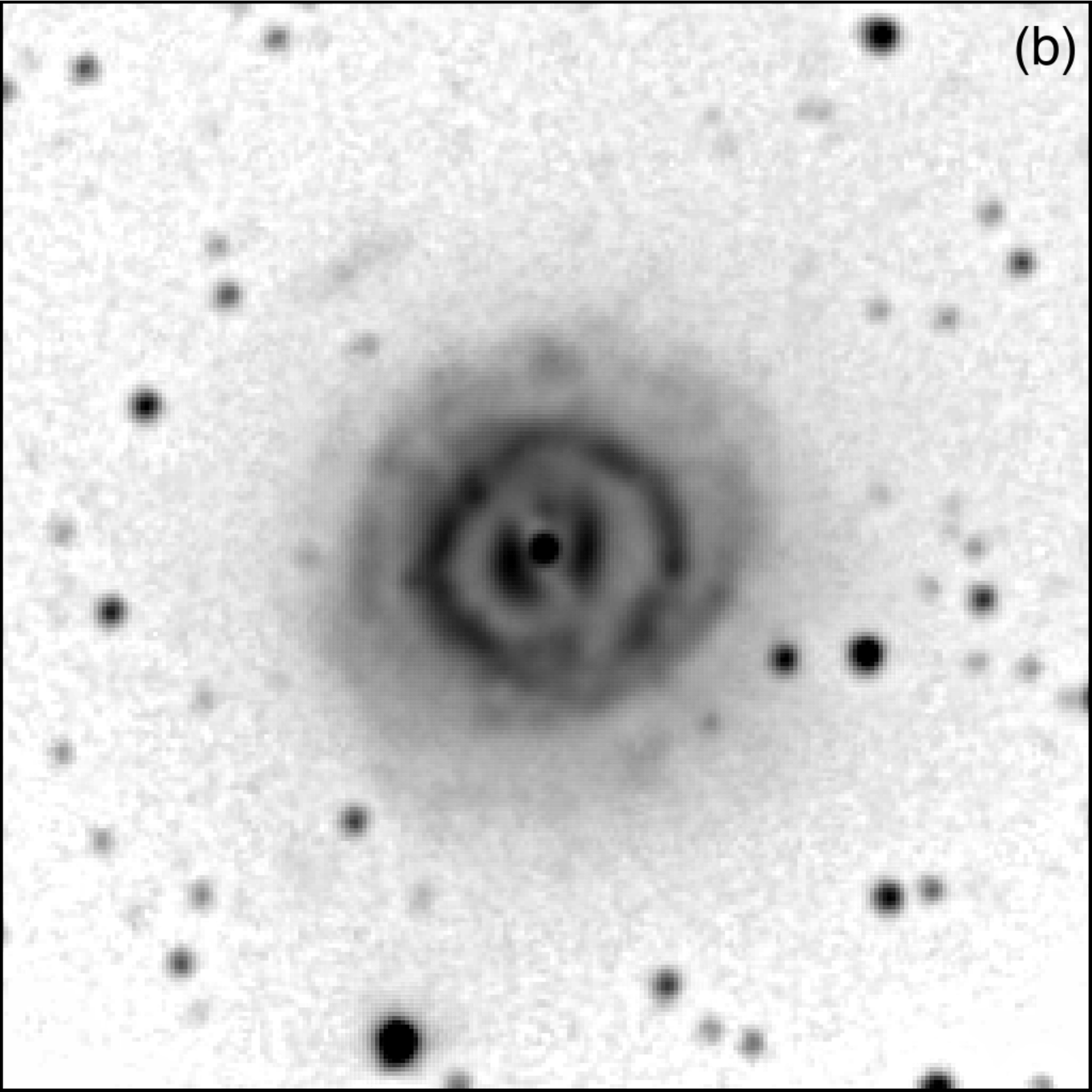}\\
      \includegraphics[scale=0.45,bb=0 0 512 512]{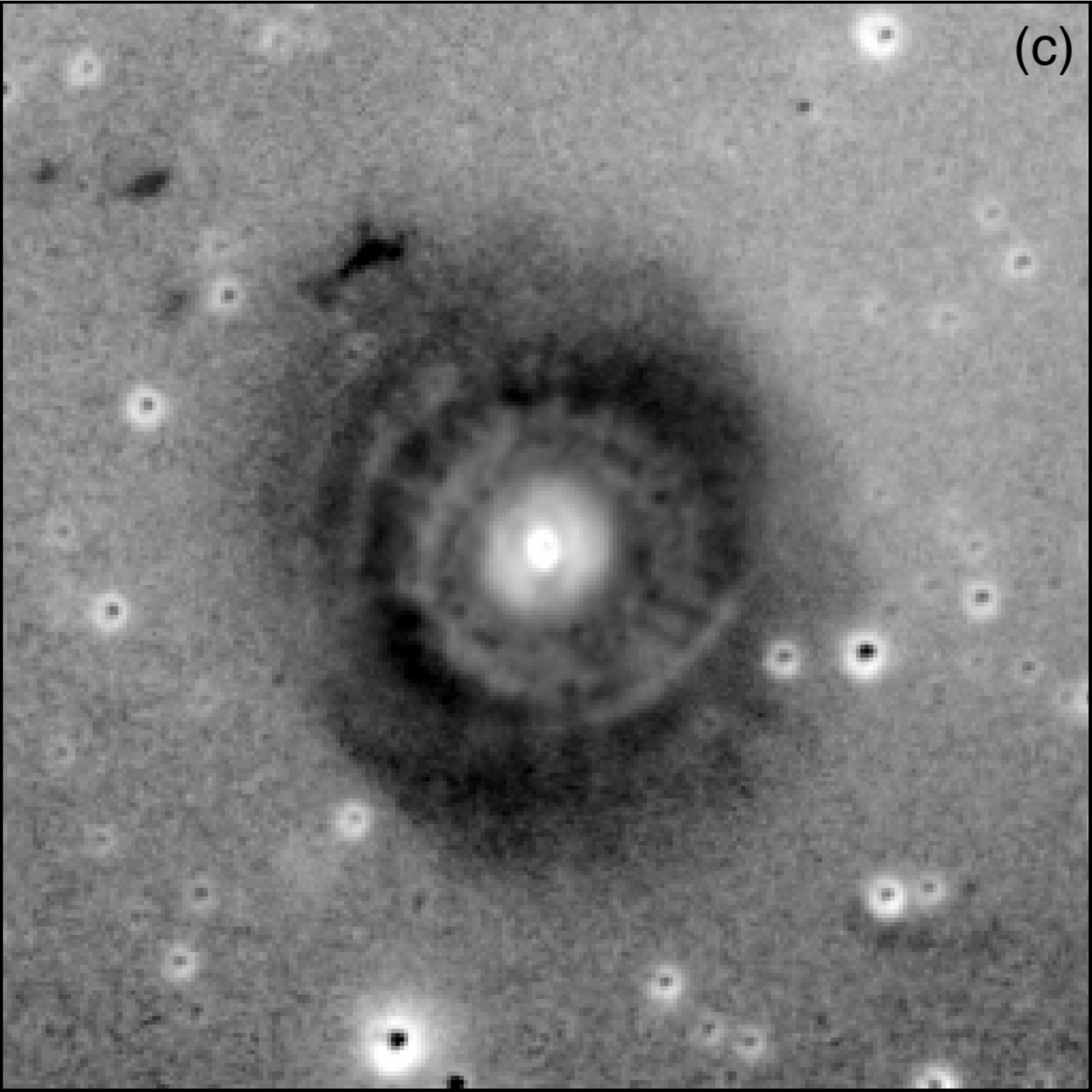}
      \includegraphics[scale=0.45,bb=0 0 512 512]{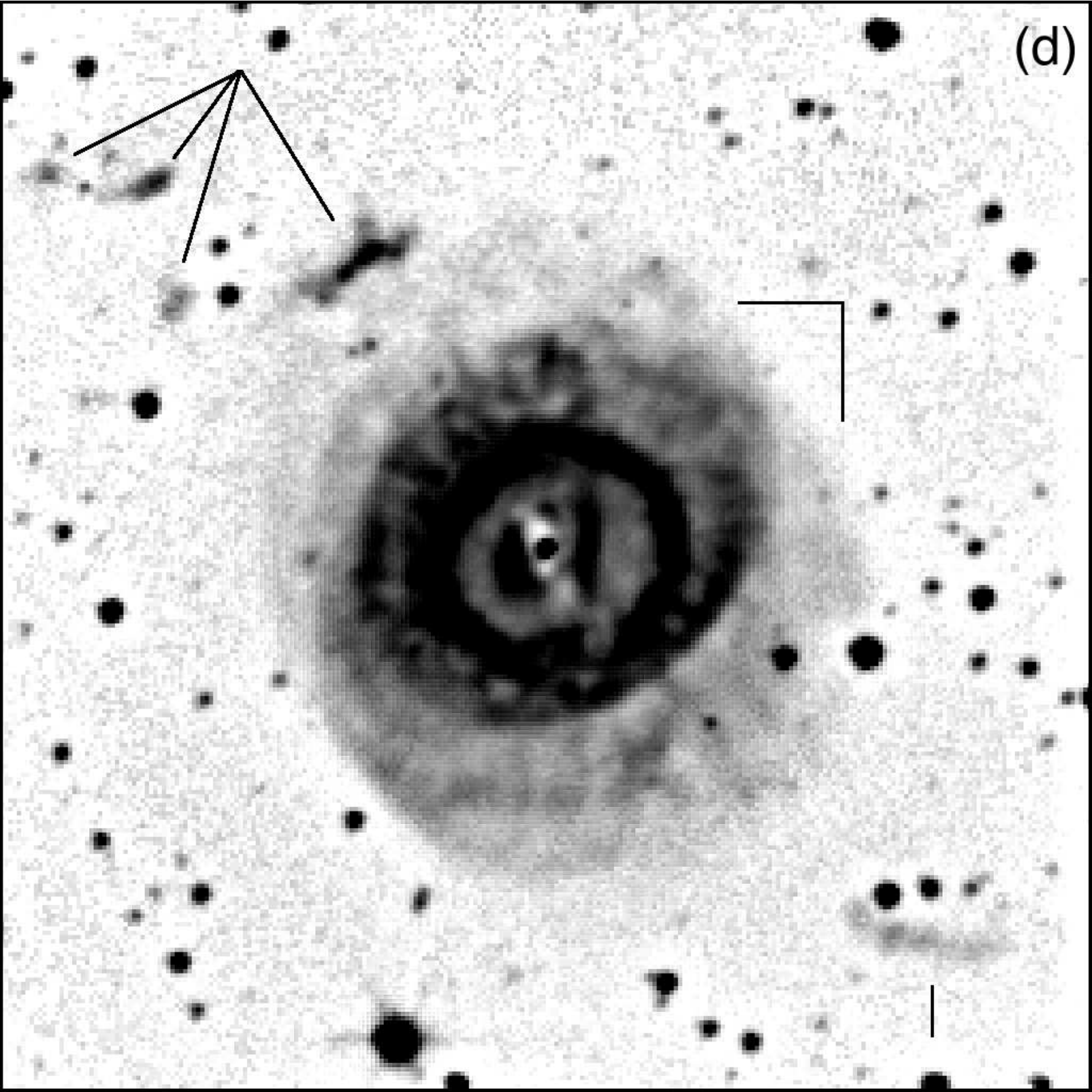}
   \end{center}
   \caption{SALT RSS Fabry-P\'erot imaging of Sp~3 in the H$\alpha$ (a) and [O~III] (b) emission lines. Panel (c) is the quotient H$\alpha$ divided by [O~III] and (d) is a version of (a) with another unsharp mask filter applied. A logarithmic scale and an unsharp mask filter was applied to all images to enhance faint features. Image dimensions are $130\times130$ arcsec$^2$ with North up and East to left. Lines in (d) indicate the positions of knots (NE and SW corners) suspected to originate from jets and bipolar lobes that are more prominent on the NW side of the nebula. Morphological features are discussed further in Sect. \ref{sec:morph}.}
   \label{fig:fp}
\end{figure*}

We used the Fabry-P\'erot imaging capability of the Robert Stobie Spectrograph (RSS; Burgh et al. 2003; Kobulnicky et al. 2003; Rangwala et al. 2008) on SALT to obtain [O~III] and H$\alpha$ images of Sp~3 on 19 September 2012 and 13 October 2012, respectively, as part of programme 2012-1-RSA\_OTH-010 (PI: Miszalski). The low-resolution etalon was tuned to the wavelength of each emission line and the contribution of [N~II] emission to the H$\alpha$ image was negligible. Images were taken in a $3\times3$ grid pattern where the telescope dithered 15\arcsec\ between each grid location. Seven [O~III] and nine H$\alpha$ exposures of 219 s each were taken in 2.05\arcsec\ and 1.45\arcsec\ seeing, respectively. Figure \ref{fig:fp} shows the final images after basic pipeline processing (Crawford et al. 2010), cosmic ray cleaning (van Dokkum 2001), aligning and median combining the data. 

\subsection{\'Echelle spectroscopy}
\begin{figure*}
   \begin{center}
      \includegraphics[scale=0.75,bb=0 0 649 670]{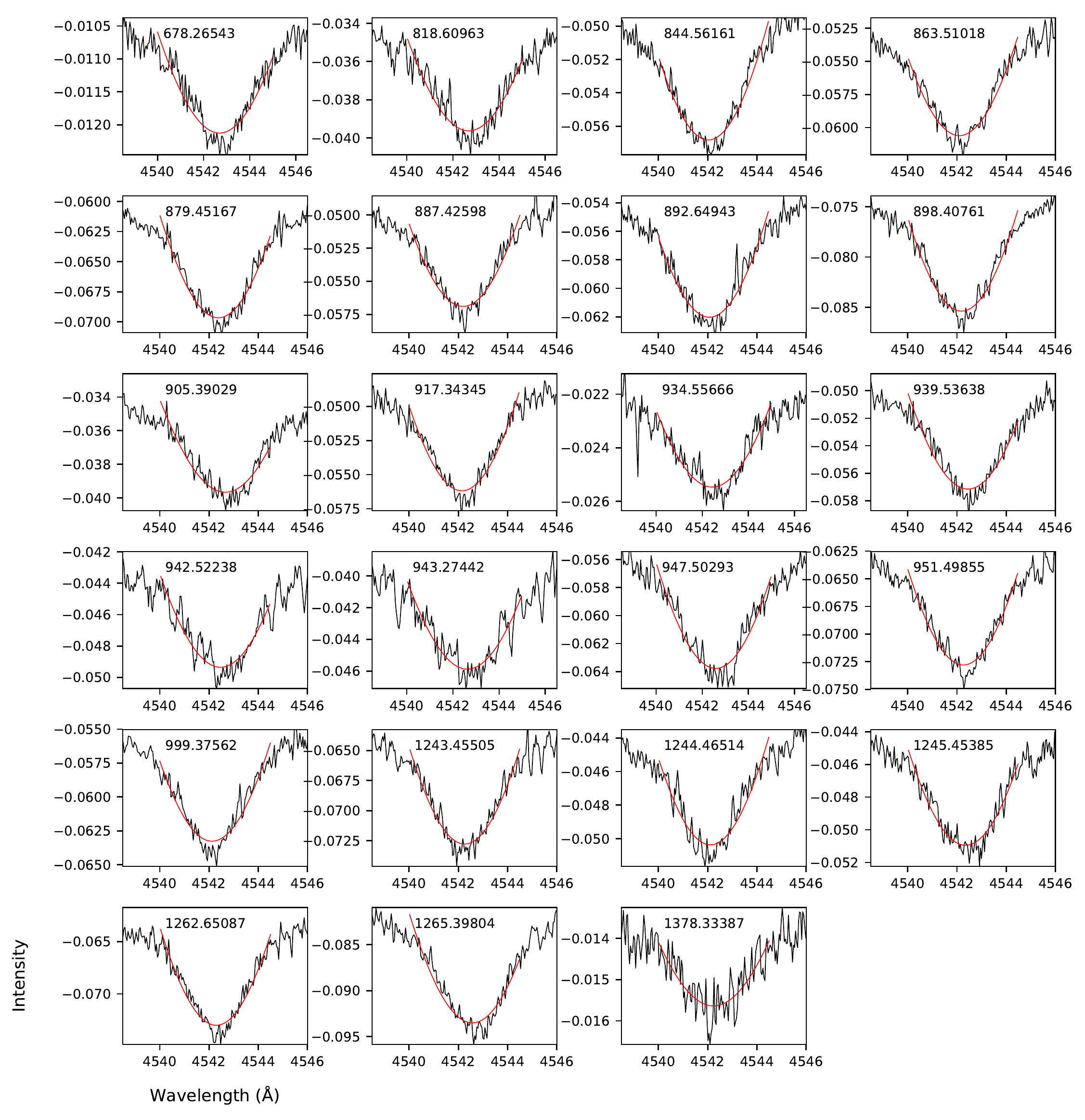}
   \end{center}
   \caption{The observed stellar He~II $\lambda$4541.59 \AA\ profiles (black lines) and the Voigt function fits (red lines). Each panel is labelled with the Julian day of each spectrum minus 2457000 days.}
   \label{fig:fitstellar}
\end{figure*}

\begin{figure*}
   \begin{center}
      \includegraphics[scale=0.75,bb=0 0 630 670]{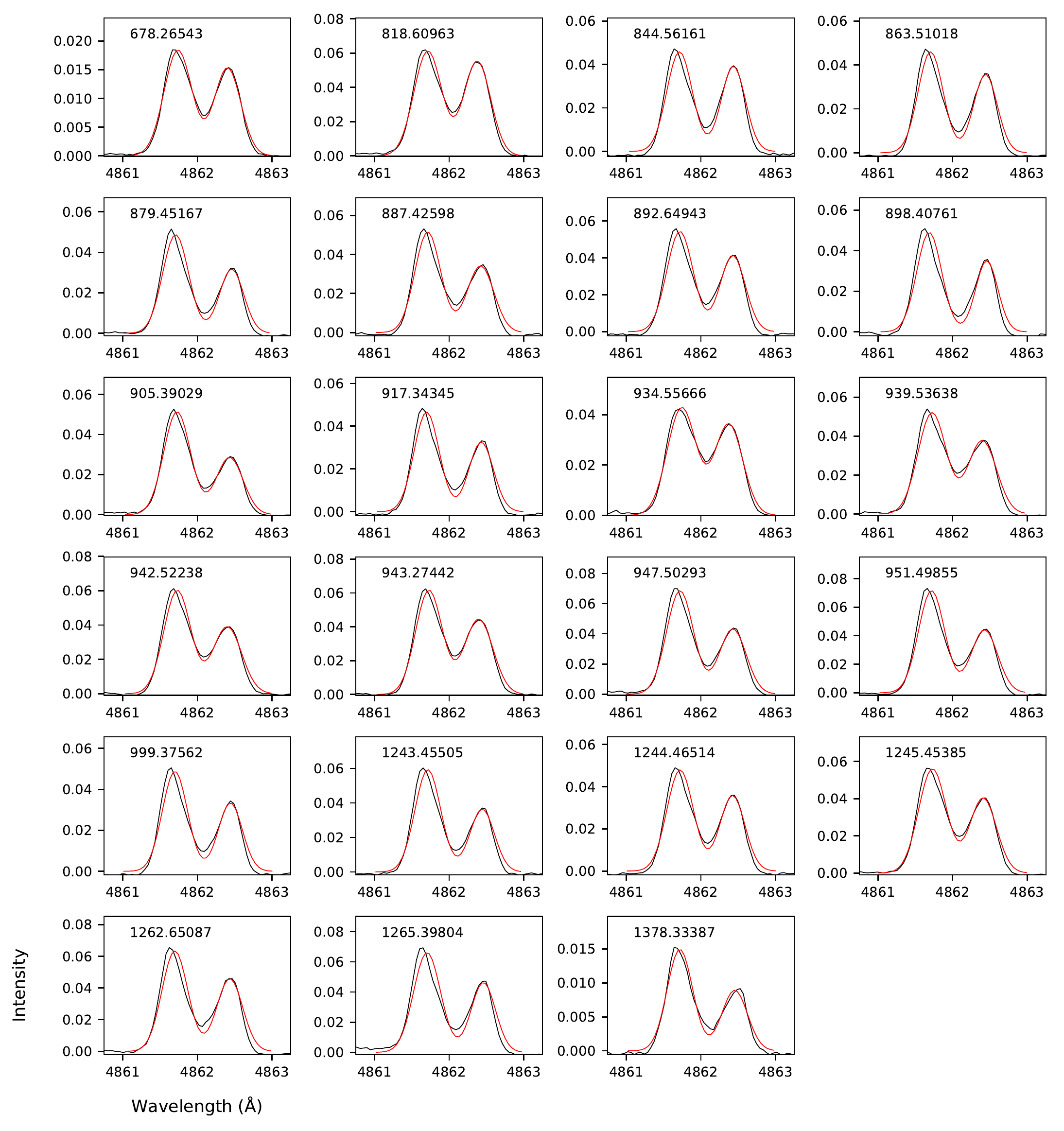}
   \end{center}
  \caption{The observed nebular H$\beta$ $\lambda$4861.363 \AA\ profiles (black lines) and the multiple Gaussian function fits (red lines). Each panel is labelled with the Julian day of each spectrum minus 2457000 days.}
   \label{fig:fitnebular}
\end{figure*}

A total of 23 \'echelle spectra of Sp~3 were obtained with the High Resolution Spectrograph (HRS) on SALT (Bramall et al. 2010, 2012; Crause et al. 2014) under programmes 2016-2-SCI-034 and 2017-1-MLT-010 (PI: Miszalski). Table \ref{tab:log} gives a log of the observations taken with the medium resolution mode. We primarily use the blue arm data (resolving power $R=\lambda/\Delta\lambda=43000$, for details see Miszalski et al. 2018a). The basic data products (Crawford et al. 2010) were reduced with the \textsc{midas} pipeline developed by Kniazev et al. (2016) which is based on the \textsc{echelle} (Ballester 1992) and \textsc{feros} (Stahl et al. 1999) packages. Heliocentric corrections were applied to the data using \textsc{velset} of the \textsc{rvsao} package (Kurtz \& Mink 1998). Radial velocity measurements in Table \ref{tab:log} were obtained by fitting single Voigt and two Gaussian functions to stellar He~II $\lambda$4540 and nebular H$\beta$ $\lambda$4861 features, respectively, using the \textsc{lmfit} package (Newville et al. 2016). Figures \ref{fig:fitstellar} and \ref{fig:fitnebular} show the fits to the data. A weighted mean of the separation of the resolved H$\beta$ emission yields an expansion velocity of $2V_\mathrm{exp}=43.1\pm0.1$ km s$^{-1}$ and a heliocentric radial velocity of $43.5\pm0.1$ km s$^{-1}$. The latter is in good agreement with $45.2\pm4.7$ km s$^{-1}$ given by Durand et al. (1998).

\begin{table}
   \centering
   \caption{Log of SALT HRS observations of Sp~3. The Julian day represents the midpoint of each exposure and radial velocity measurements are made from stellar He~II $\lambda$4540 and nebular H$\beta$ 4861.}
   \label{tab:log}
   \begin{tabular}{lrll}
      \hline\hline
      Julian day & Exposure & RV (He~II) & RV (H$\beta$)\\
                 & time (s) &  (km s$^{-1}$) & (km s$^{-1}$)\\
      \hline
      2457678.26543 & 2250 & 76.02$\pm$1.50&43.72$\pm$0.18 \\
      2457818.60963 & 2050 & 77.12$\pm$1.87&42.09$\pm$0.17 \\
      2457844.56161 & 2050 & 26.74$\pm$1.05&43.87$\pm$0.25 \\
      2457863.51018 & 2050 & 32.98$\pm$1.20&43.73$\pm$0.24 \\
      2457879.45167 & 2050 & 54.50$\pm$1.04&44.44$\pm$0.26 \\
      2457887.42598 & 2050 & 39.38$\pm$1.17&43.42$\pm$0.31 \\
      2457892.64943 & 2050 & 29.80$\pm$1.35&43.75$\pm$0.26 \\
      2457898.40761 & 2050 & 38.94$\pm$0.99&43.71$\pm$0.27 \\
      2457905.39029 & 2050 & 74.90$\pm$1.61&44.20$\pm$0.22 \\
      2457917.34345 & 2050 & 36.79$\pm$1.01&43.23$\pm$0.28 \\
      2457934.55666 & 2050 & 56.85$\pm$1.92&42.87$\pm$0.20 \\
      2457939.53638 & 2050 & 59.62$\pm$1.26&43.02$\pm$0.28 \\
      2457942.52238 & 2050 & 62.74$\pm$2.01&43.32$\pm$0.25 \\
      2457943.27442 & 2050 & 71.77$\pm$2.39&43.13$\pm$0.21 \\
      2457947.50293 & 2050 & 67.68$\pm$1.35&43.83$\pm$0.22 \\
      2457951.49855 & 2050 & 45.53$\pm$1.06&43.68$\pm$0.25 \\
      2457999.37562 & 2050 & 33.87$\pm$1.14&43.64$\pm$0.28 \\
      2458243.45505 & 2050 & 43.75$\pm$1.05&44.15$\pm$0.26 \\
      2458244.46514 & 2050 & 32.08$\pm$1.42&43.73$\pm$0.25 \\
      2458245.45385 & 2050 & 52.50$\pm$1.44&43.52$\pm$0.23 \\
      2458262.65087 & 2050 & 47.90$\pm$0.96&43.36$\pm$0.27 \\
      2458265.39804 & 2050 & 69.04$\pm$1.13&44.31$\pm$0.27 \\
      2458378.33387 & 2050 & 42.52$\pm$3.43&44.56$\pm$0.30 \\
      \hline
   \end{tabular}
\end{table}

\subsection{Longslit spectroscopy}
Longslit observations of Sp~3 were also conducted with RSS (Burgh et al. 2003; Kobulnicky et al. 2003) on 11 June 2018 under programme 2017-1-MLT-010 to measure the chemical abundances of the nebula. The 1.25\arcsec\ wide longslit was centred on the central star with a position angle (PA) of 104 deg to place the inner [O~III] lobes near the central star on the slit (see Fig. \ref{fig:fp}b and Sect. \ref{sec:morph}). During the SALT track, exposures of 180 s and 1500 s were taken with the PG900 grating configured to cover 4350--7405 \AA. This was then followed by a 1500 s exposure taken with the PG2300 grating configured to cover 3693--4776 \AA. The exposures were binned $2\times2$ before read out and the resulting approximate spectral resolutions measured from arc lamp emission lines were 4.80 and 1.75 \AA, respectively. After basic reductions were performed by \textsc{pysalt} (Crawford et al. 2010), cosmic ray events were cleaned using the \textsc{lacosmic} package (van Dokkum 2001) before the data were reduced using standard \textsc{iraf} routines such as \textsc{identify}, \textsc{reidentify}, \textsc{fitcoords} and \textsc{transform}.

\begin{figure}
   \begin{center}
      \includegraphics[scale=0.5,bb=0 0 347 347]{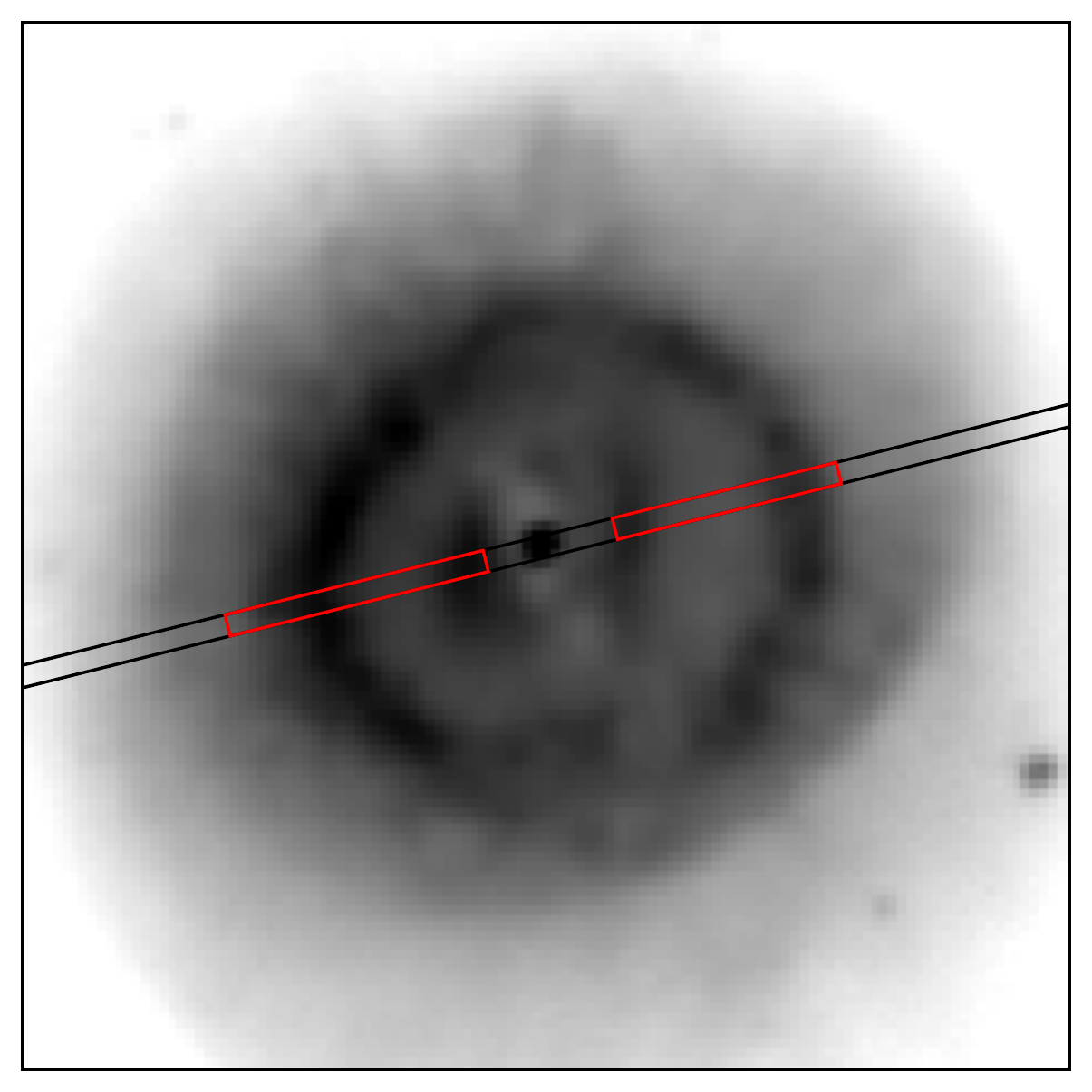}
      \includegraphics[scale=0.5,bb=0 0 414 176]{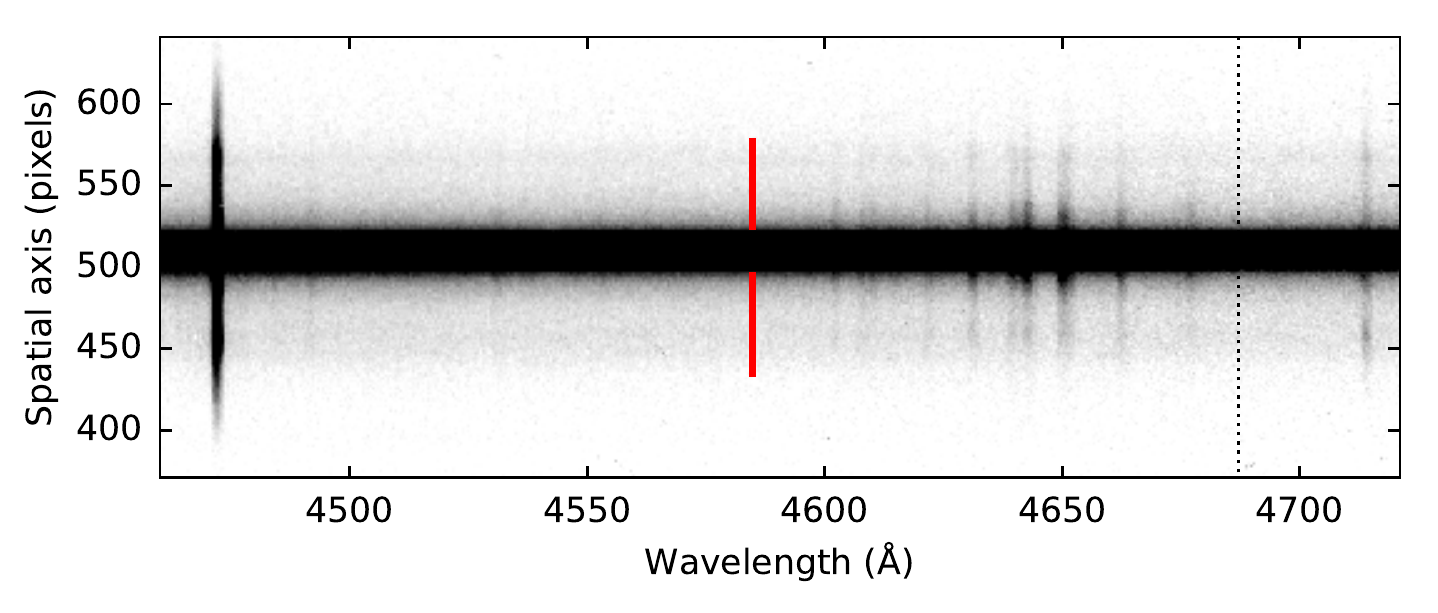}
   \end{center}
   \caption{\emph{(Top panel)} Position of the RSS 1.25\arcsec\ longslit with a PA of 104 deg (black rectangle) on the H$\alpha$ image of Sp~3 (Fig. \ref{fig:fp}). Red rectangles of 15.2\arcsec\ (left) and 13.2\arcsec\ (right) indicate the apertures used to extract integrated spectra. Image dimensions are $60\times60$ arcsec$^2$ and the orientation is the same as Fig. \ref{fig:fp}. (\emph{Bottom panel}) Part of the PG2300 spectrum showing the nebular nature of the recombination lines near 4650 \AA. The dotted line indicates the expected location of the undetected He~II $\lambda$4686 emission line. The same apertures as in the \emph{top panel} are indicated by red lines either side of the central star. The spatial scale is 0.254\arcsec\ per pixel.}
   \label{fig:slit}
\end{figure}

\begin{figure*}
   \begin{center}
      \includegraphics[scale=0.45,bb=0 0 1073 494]{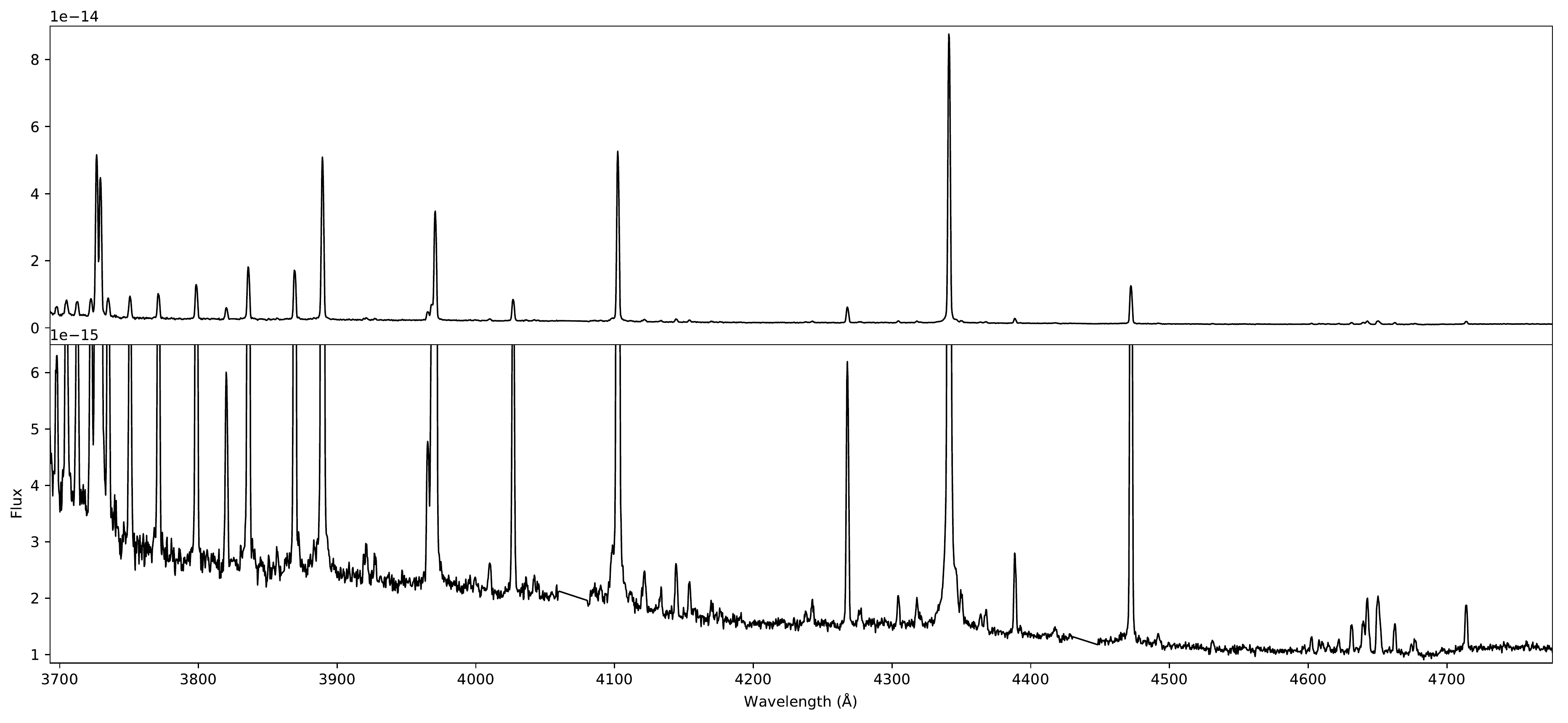}
      \includegraphics[scale=0.45,bb=0 0 1073 494]{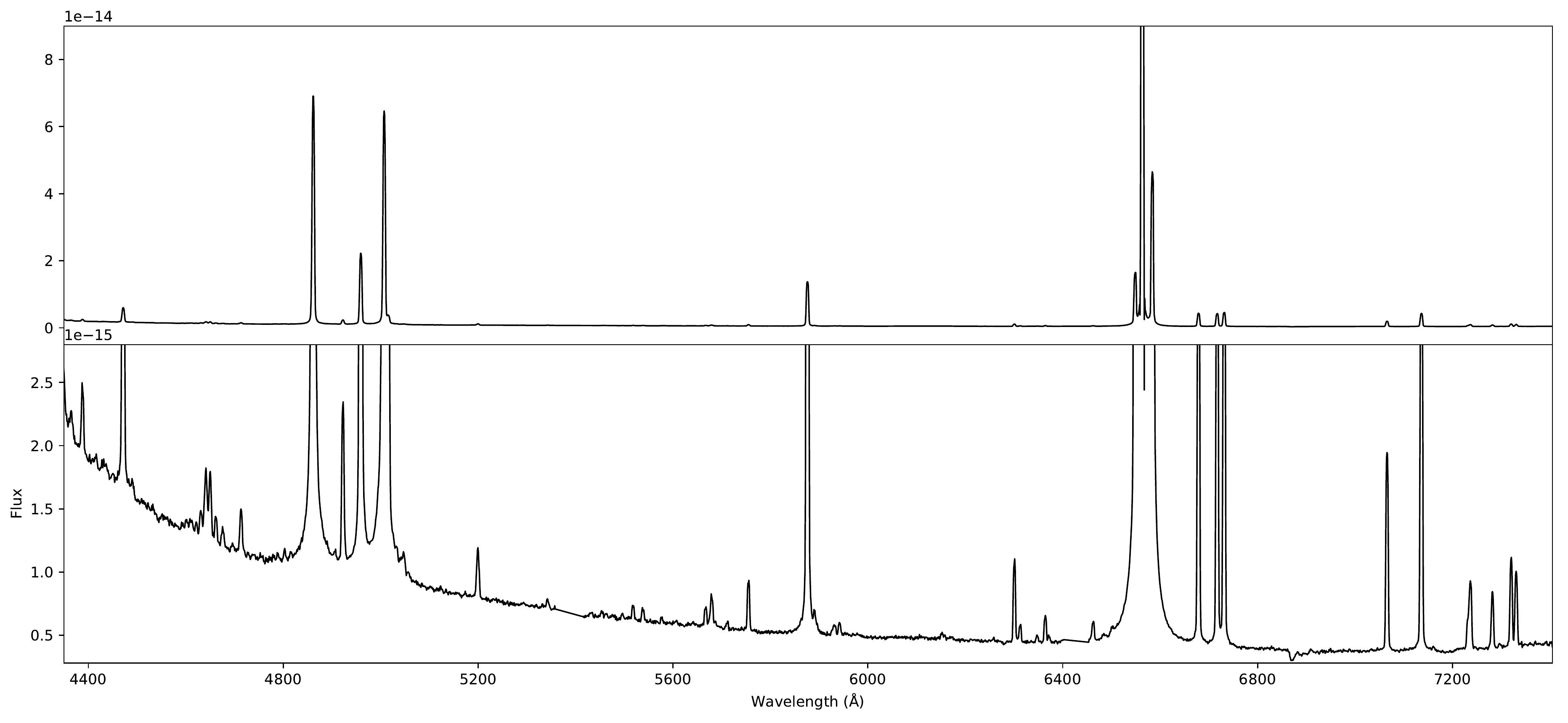}
   \end{center}
   \caption{The average integrated RSS spectra extracted from the exposures taken with the PG2300 (top) and PG900 (bottom) gratings. Line identifications are listed in Table \ref{tab:lines1}.} 
   \label{fig:spectra}
\end{figure*}

The PG2300 spectrum clearly showed optical recombination lines visible in the brightest inner part of the nebula (Fig. \ref{fig:slit}). Figure \ref{fig:slit} shows the two windows either side of the central star that were used to extract integrated spectra for chemical abundance analysis (Sect. \ref{sec:chem}) and the sky background was subtracted from regions well outside the whole nebula. The \textsc{iraf} task \textsc{apall} was used to extract spectra from these windows before being averaged into a single spectrum per observation. The same window was extracted from PG2300 and PG900 spectra relative to the trace of the central star determined by \textsc{apall}. Figure \ref{fig:spectra} shows the average spectra which were flux calibrated using spectra of the spectrophotometric standard stars EG274 (PG900) and G93-48 (PG2300). The absolute value of the flux calibration should only be considered to be approximate due to the moving pupil design of SALT. A separate spectrum of the central star was extracted and used to check that the relative calibration is smooth across both spectra, including in the overlap region, and that no additional features were imprinted onto the spectra due to flux calibration. 

\section{ANALYSIS}
\label{sec:analysis}
\subsection{Nebular morphology}
\label{sec:morph}
Figure \ref{fig:fp} reveals new morphological details not evident in previous images (Schwarz et al. 1992). Faint outer lobes appear to emerge in the H$\alpha$ image from a minor axis with a PA of $\sim$125 deg. These lobes do not appear in the [O~III] image and are most prominent on the NW side of the nebula (Fig. \ref{fig:fp}d). The outer lobes suggest the underlying morphology is bipolar. We measure a nebula radius of $\sim$34\arcsec\ from a contour based on 10 per cent of the average H$\alpha$ brightness in the inner nebula. The brightest features are an apparently broken ring of radius 14\arcsec\ and an inner pair of lobes that is brightest in [O~III]. These inner lobes are visible in Fig. \ref{fig:fp}b near the central star and are brighter on the E and W sides of the central star. They are reminiscent of the inner [O~III] emission observed in the bipolar post-CE PN M2-19 (Miszalski et al. 2009b). Several faint knots are located outside the main nebula with four to the NE and one to the SW (Fig. \ref{fig:fp}d). Their appearance is similar to jets in the post-CE PN NGC~6337 which is viewed almost pole-on to the line-of-sight (e.g. NGC~6337, Garc\'ia-D\'iaz et al. 2009). A thorough spatiokinematic study of the nebula is encouraged to further investigate its unusual morphology, jet system and inclination angle. We discuss possible inclination angles further in Sect. \ref{sec:orb}. 

\subsection{Photospheric parameters and mass of the primary}
\label{sec:atmos}
Gauba et al. (2001) examined low-resolution ($R = \lambda/\Delta\lambda \approx 300$) ultraviolet (UV) spectra of the central star of Sp~3 obtained with the \emph{International Ultraviolet Explorer (IUE)}. They determined an O3V spectral type with an effective temperature of about \Teffw{50\,000} by comparison with the spectrophotometric standard star HD 93205. From the P-Cygni profile of the C~IV $\lambda\lambda$ 1548, 1551 \AA\, resonance lines, they also measured a terminal wind velocity $v_\infty = 1603 \pm 400$ km s$^{-1}$. From the presence of a stellar wind, they concluded that the surface gravity is $\log g < 5.2$  cm s$^{-2}$ (Cerruti-Sola \& Perinotto 1985). Guerrero \& De Marco (2013) found variability in the UV spectra, but not enough epochs were available to identify its cause. 

To determine the stellar parameters of the primary, we corrected several individual orders of the blue HRS spectra for orbital motion (Tab. \ref{tab:log}) and created average spectra around some strategic absorption lines that are suited for a detailed spectral analysis. Since non-local thermodynamic equilibrium (NLTE) atmosphere models are mandatory for such a hot star (e.g. Rauch et al. 2018), we employed the T\"ubingen non-LTE Model-Atmosphere Package (TMAP\footnote{https://uni-tuebingen.de/de/41621}; Werner et al. 2003, 2012; Rauch \& Deetjen 2003) to calculate plane-parallel models in radiative and hydrostatic equilibrium.

\begin{figure}
   \includegraphics[scale=1.0,bb=0 0 233 245]{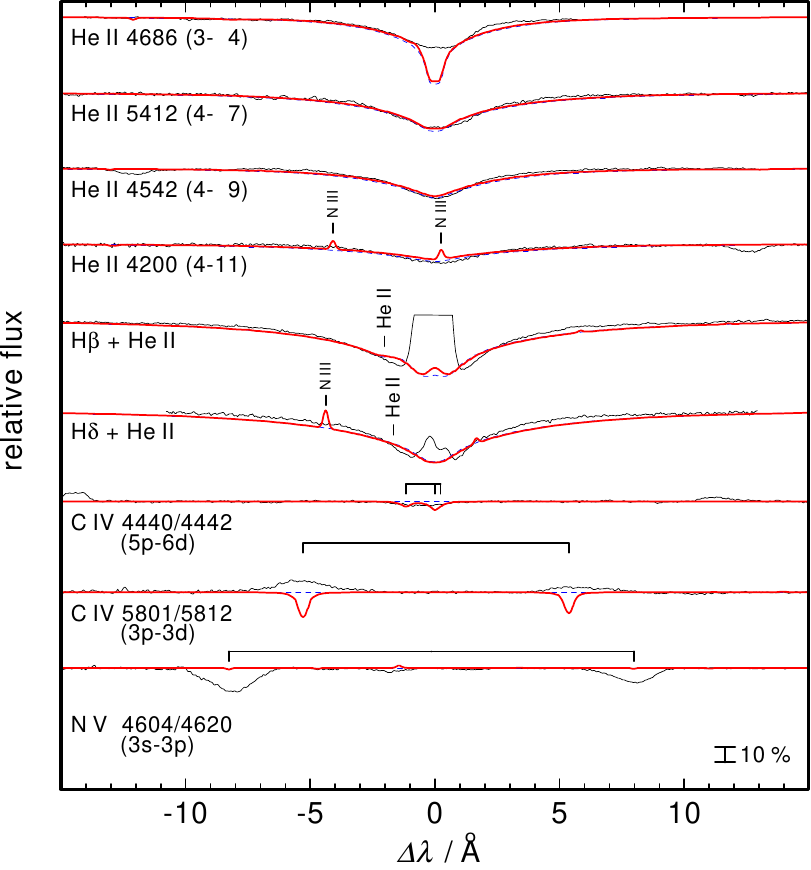}
    \caption{Sections of the HRS spectra (black) compared with two synthetic spectra from models with \Teffw{50\,000} and \loggw{5.0} composed of He+He (blue, dashed line) and H+He+C+N (red, thick line). All abundances are solar. All spectra shown were convolved with Gaussians according to the HRS spectral resolution. }
   \label{fig:5050}
\end{figure}

Since lines of H, He, C, and N are prominent in the observed spectra, we calculated two models
composed of H+He and H+He+C+N with solar abundances adopted from Asplund et al. (2009) 
with \Teffw{50\,000} and \loggw{5.0} (Fig.\,\ref{fig:5050}). For the H~I and He~II lines, we find a good agreement between these models. The outer line wings of the H\,$\beta$/He~II and H\,$\delta$/He~II lines are too strong compared with the observed profiles, indicating a lower \logg.

\begin{figure}
   \includegraphics[scale=1.0,bb=0 0 233 165]{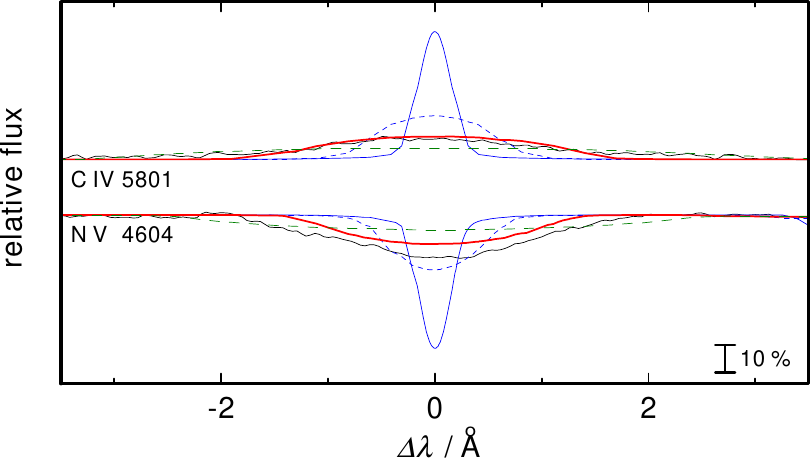}
   \caption{Comparison of the observed profiles of C~IV $\lambda$5801.31 \AA\, and N~V $\lambda$4603.74 \AA\,(black lines) with profiles calculated from a model with \Teffw{68\,000} and \loggw{4.6}. The synthetic spectra are convolved with a rotational profile with $v_\mathrm{rot} =$ 0 (blue, thin line), 40 (blue, dashed line), 80 (red line), and 120\,km s$^{-1}$ (green, dashed line). The C and N mass fractions were adjusted to match the equivalent widths of the observed line profiles at $v_\mathrm{rot} = 80$\,km s$^{-1}$ in this figure.}
   \label{fig:rotation}
\end{figure}

We calculated an extended grid of NLTE model atmospheres within
$50\,000\,\mathrm{K}\, \sla\, \Teff\, \sla\, 82\,000\,\mathrm{K}$ (with steps of 2\,000\,K),
and $4.5\, \sla\, \logg\, \sla\, 5.0 $ (0.1) that consider opacities of H+He+C+N with solar abundances. While the outer lines wings of the H\,$\beta$/He~II and H\,$\delta$/He~II blends are well reproduced at \loggw{4.6 \pm 0.2}, the theoretical line profiles of N~V $\lambda\lambda$ 4604, 4620 \AA\, and C~IV $\lambda\lambda$ 5801, 5812 \AA\, are much too narrow to reproduce the observed profiles. A significant rotation of $v_\mathrm{rot} \approx 80$ km s$^{-1}$ is necessary for a reasonable fit (Fig.\,\ref{fig:rotation}) and we adopt this value for our further analysis.

\begin{figure}
   \includegraphics[scale=1.0,bb=0 0 233 250]{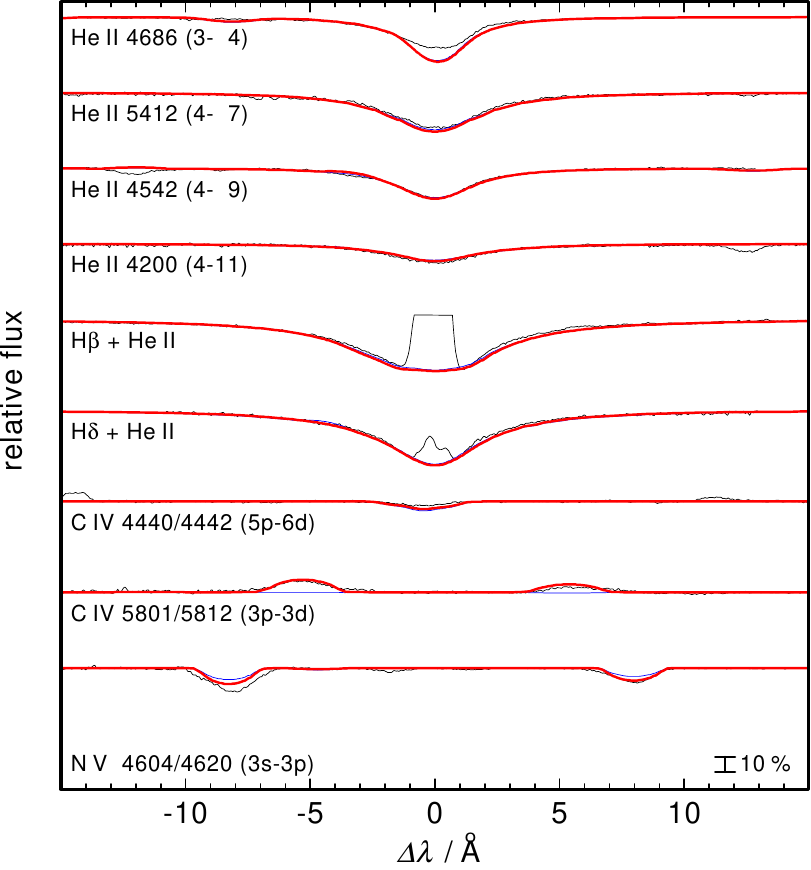}
   \caption{Same as in Fig.\,\ref{fig:5050}, but for two models with \Teffw{60\,000} (blue, thin line) and  \Teffw{68\,000} (red, thick line), \loggw{4.6}, [H] = 0.05, [He] = 0.02, [C] = $-$0.088, and [N] = 0.39.  [X] denotes log(fraction of element X / solar fraction of X). The synthetic spectra consider $v_\mathrm{rot} = 80$ km s$^{-1}$.}
   \label{fig:6846}
\end{figure}

The determination of \Teff\ is hampered because no lines of subsequent ionization stages of one element could be identified in the available spectra to evaluate its ionization equilibrium precisely. However, we found that in general, C~IV $\lambda\lambda$ 5801, 5812 \AA\, turns into emission only for \Teff $> 60\,000\,\mathrm{K}$, while C~IV $\lambda\lambda$ 4440, 4442 \AA\, remains in absorption. Figure\,\ref{fig:6846} shows a comparison of a model with \Teffw{68\,000} and \loggw{4.6} to the observed spectra. All theoretical line profiles are in good agreement with the observations, but He~II $\lambda$4686.06 \AA\, is much shallower than expected.

\begin{figure}
   \includegraphics[scale=1.0,bb=0 0 233 131]{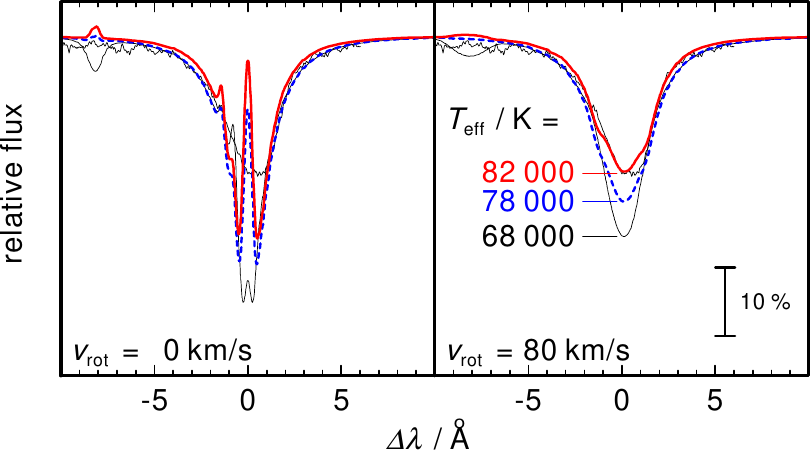}
   \caption{The HRS spectrum around He~II $\lambda$4686.06 \AA\, compared to three models with \loggw{4.6} and different \Teff for $v_\mathrm{rot}=0$ km s$^{-1}$ (left) and $v_\mathrm{rot}=80$ km s$^{-1}$ (right).}
   \label{fig:teff}
\end{figure}

\begin{figure}
   \includegraphics[scale=1.0,bb=0 0 233 103]{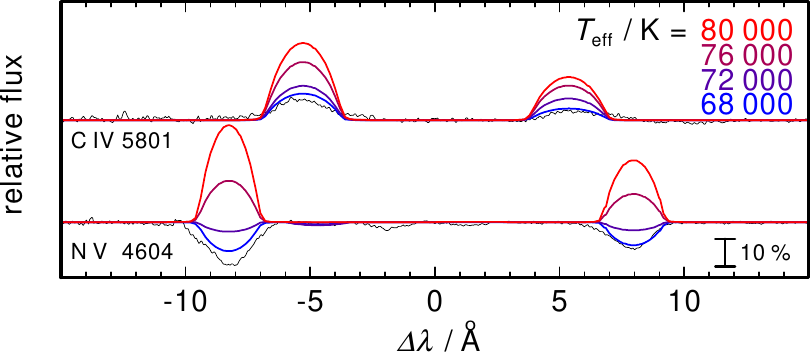}
    \caption{Sections of the HRS spectrum (black) compared with synthetic spectra from models with         
             \Teff\ = 68\,000, 72\,000, 76\,000, and 80\,000 K,
             \loggw{4.6},
             [H]  = 0.05,
             [He] = 0.02,
             [C]  = $-$0.088, and
             [N]  = 0.39.
            }
   \label{fig:abtoem}
\end{figure}

The observed He~II $\lambda$4686.06 \AA\,  line profile is obviously asymmetric, most likely due to 
problems in the data reduction. It is located at the red end of an HRS \'echelle order and the 
rectification of the outer red line wing is therefore difficult. The central depression, however, 
should not be affected significantly and this was confirmed independently by comparing the HRS spectrum with the PG900 RSS spectrum of the central star (Sect. \ref{sec:obs}). Rauch et al. (1996) have shown that due to a temperature inversion in the photosphere, an emission reversal in the line center of He~II $\lambda$4686.06 \AA\,  is a sensitive indicator of \Teff\ because it strengthens with increasing \Teff. The rapid stellar rotation is then responsible for a shallower line core at higher \Teff. Figure\,\ref{fig:teff} demonstrates this effect where we can reproduce the observed He~II $\lambda$4686.06 \AA\, with a \Teffw{82\,000} and \loggw{4.6} model. From N~V $\lambda\lambda$4604, 4620 \AA, we have an additional constraint because it turns into emission for \Teff $\sga 74\,000$\,K (Fig.\,\ref{fig:abtoem}). Furthermore, the H~I/He~II blends become deeper than observed for \Teff $\sga 70\,000$\,K. Thus, we adopt \Teffw{68\,000^{+12\,000}_{-6\,000}}.

Figure\,\ref{fig:evolution} shows the CSPN of Sp~3 in the $\log$ \Teff\,--\,\logg\ diagram compared to stellar evolutionary tracks of H-rich post-AGB stars (Miller Bertolami et al. 2016). We interpolate from these tracks a stellar mass of $M = 0.60^{+0.27}_{-0.05}$\,\Msol. From the tables of Miller Bertolami et al. (2016) we determine a stellar luminosity of $\log (L\,/\,L_\odot) = 3.85^{+0.55}_{-0.35}$. The position of Sp~3 in Fig. \ref{fig:evolution} is consistent with a post-AGB origin, assuming these single star tracks are applicable to the binary central star, rather than the post-RGB origin suggested by Hillwig et al. (2017). The location of the CSPN of Sp~3 is relatively close to the Eddington limit (Fig.\,\ref{fig:evolution}) and, thus, mass loss due to the stellar wind may have an impact on the spectral analysis, especially on the strengths of the C~IV $\lambda\lambda$ 5801, 5812 \AA\, emission lines. Table \ref{tab:results} summarises the results of our TMAP NLTE analysis. 

\begin{figure}
   \includegraphics[scale=0.9]{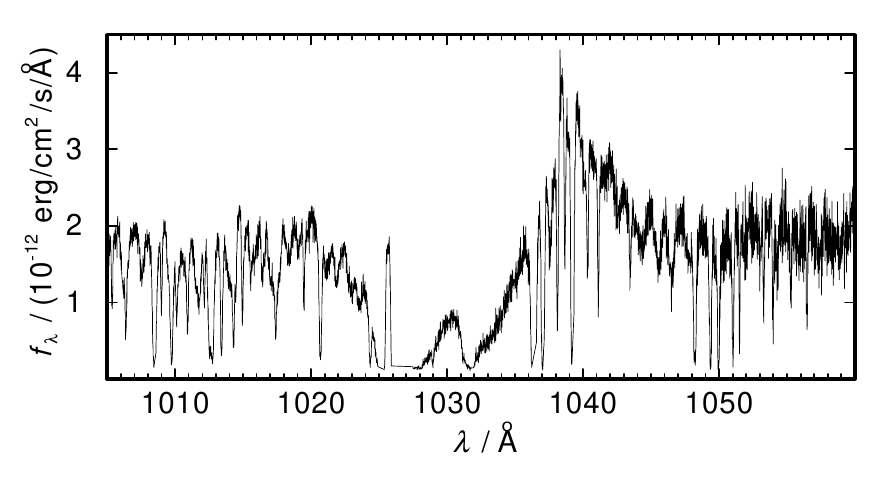}
   \caption{Section of the \emph{FUSE} observation around O~VI $\lambda\lambda$ 1032,1038 \AA.}
    \label{fig:fuse}
\end{figure}

\begin{figure}
   \includegraphics[scale=0.5,bb=0 0 473 450]{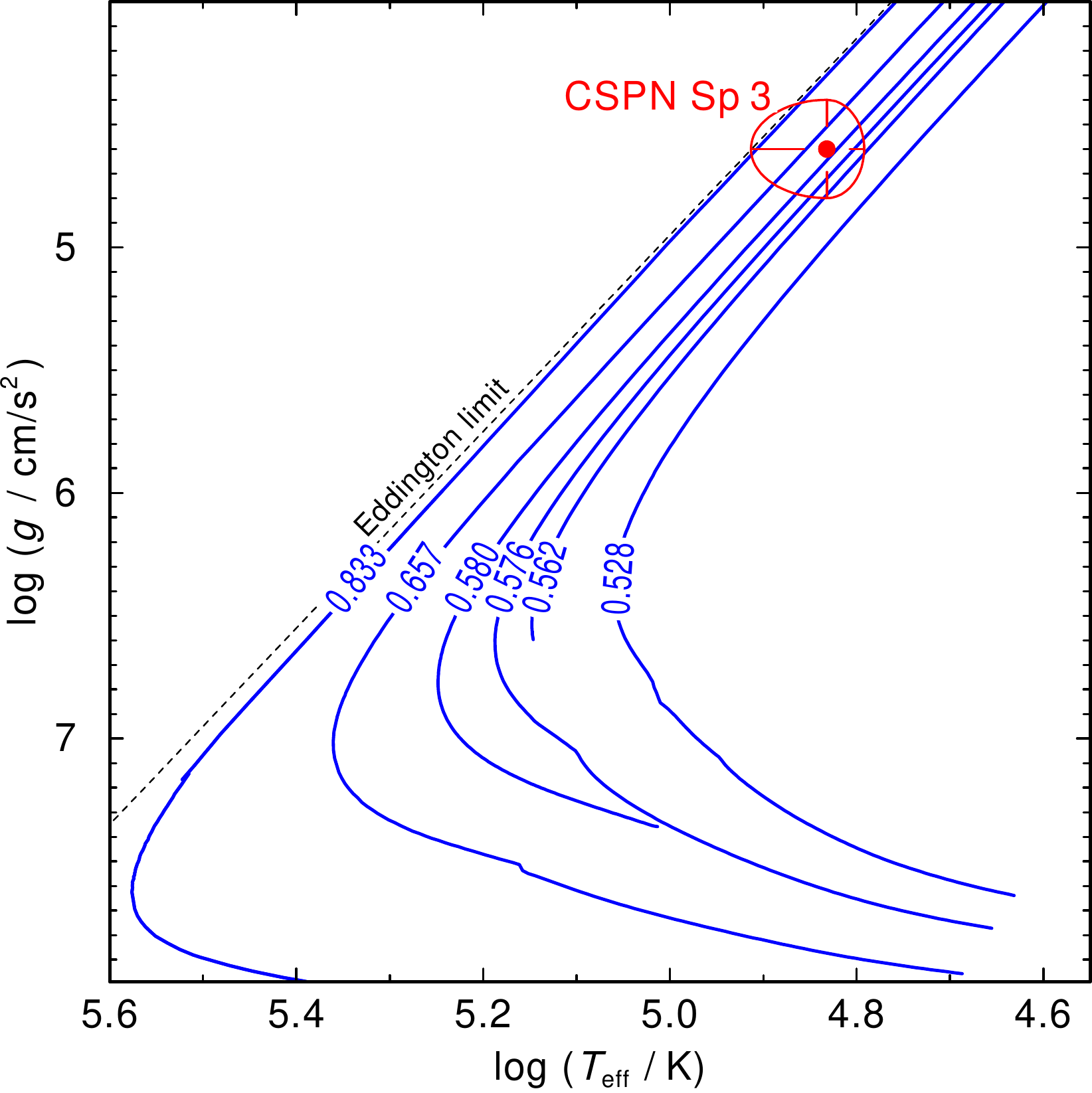}
    \caption{Location of the CSPN of Sp~3 (with its error range) in the $\log$ \Teff\,--\,\logg\ plane. Post-AGB evolutionary tracks of H-rich stars (for about solar metallicity, $Z = 0.02$; Miller Bertolami et al. 2016) labeled with the stellar mass in $M_\odot$, respectively, are shown for comparison. The dashed, black line indicates the Eddington limit for solar abundances.}
   \label{fig:evolution}
\end{figure}

To improve the spectral analysis, high-resolution UV spectroscopy with a high signal-to-noise ratio is highly desirable to investigate the wind properties and to determine \Teff\ based  on multiple ionization equilibria of metal lines that form in the static region of the photosphere. Unfortunately, an available \emph{FUSE}\footnote{Far Ultraviolet Spectroscopic Explorer.} far-UV observation is strongly contaminated by interstellar line absorption and is thus not suitable for a precise spectral analysis. However, a P-Cygni profile of the O~VI $\lambda\lambda$ 1032,1038 \AA\ resonance doublet is prominent in the \emph{FUSE} observation (Id B032080100000, LWRS aperture, 9439\,s exposure time, TTAG mode, Fig.\,\ref{fig:fuse}), as expected from the presence of the C~IV $\lambda\lambda$ 1548,1551 \AA\ P-Cygni profile in the IUE spectra (Gauba et al. 2001). A detailed re-analysis of the wind properties is beyond the scope of this paper.

\begin{table}\centering
  \caption{Parameters of the CSPN of Sp~3 as derived by our TMAP NLTE analysis.}         
\label{tab:results}
\setlength{\tabcolsep}{.4em}
\begin{tabular}{rlr@{.}lr@{.}lr@{.}l}
\hline
\hline
\noalign{\smallskip}                                                                                                     
\multicolumn{2}{l}{$T_\mathrm{eff}$ (K)}             & \multicolumn{2}{l}{$68\,000^{+12\,000}_{\,\,\,-6\,000}$} & \multicolumn{4}{l}{}    \\
\multicolumn{2}{l}{\logg (cm s$^{-2}$)} & \multicolumn{2}{l}{$4.6\pm 0.2$}                 & \multicolumn{4}{l}{} \\
\hline
\noalign{\smallskip}                                                                                          
&                     & \multicolumn{2}{c}{mass} & \multicolumn{2}{c}{number} & \multicolumn{2}{c}{}                      \\
\cline{3-6}                     
\multicolumn{8}{c}{}                                                                                         \vspace{-5mm}\\
& element             & \multicolumn{2}{c}{}     & \multicolumn{2}{c}{}       & \multicolumn{2}{c}{~~~~~[X]} \vspace{-2mm}\\
&                     & \multicolumn{4}{c}{fraction}                          & \multicolumn{2}{c}{}                      \\
\cline{2-8}                     
\noalign{\smallskip}                                                                                   
\smspr & \mmspr H  & $ 7$&$5\times 10^{-1}$ & $ 9$&$2\times 10^{-1}$ & $  0$&$005$ \\
       & \mmspr He & $ 2$&$5\times 10^{-1}$ & $ 7$&$8\times 10^{-1}$ & $  0$&$002$ \\
       & \mmspr C  & $ 1$&$9\times 10^{-3}$ & $ 2$&$0\times 10^{-4}$ & $ -0$&$088$ \\
       & \mmspr N  & $ 1$&$7\times 10^{-3}$ & $ 1$&$5\times 10^{-4}$ & $  0$&$387$ \\
\hline
\noalign{\smallskip}                                                                                          
\multicolumn{2}{l}{$v_\mathrm{rot}$ (km s$^{-1}$)}  & \multicolumn{2}{l}{$80 \pm 20$}          & \multicolumn{4}{l}{} \\
\multicolumn{2}{l}{\ebv\,(mag)}                               & \multicolumn{2}{l}{$0.14 \pm 0.05$}      & \multicolumn{4}{l}{} \\
\noalign{\smallskip}                                                                                          
\multicolumn{2}{l}{$M$ ($M_\odot$)}                     & \multicolumn{2}{l}{$0.60^{+0.27}_{-0.05}$} & \multicolumn{4}{l}{} \\
\noalign{\smallskip}                                                                                     
\multicolumn{2}{l}{$\log\ ( L\,/\,L_\odot )$}           & \multicolumn{2}{l}{$3.85^{+0.55}_{-0.35}$} & \multicolumn{4}{l}{} \\
\noalign{\smallskip}
\hline         
\hline
\end{tabular}\vspace{1mm}\newline
{\footnotesize
\noindent
Notes: The abundance uncertainties are estimated to be $\pm 0.5$ dex 
       (including the error propagation from the \Teff\ and \logg uncertainties).}
\end{table}

\subsection{Orbital parameters}
\label{sec:orb}
\begin{figure}
   \begin{center}
      \includegraphics[scale=0.45,bb=0 0 576 432]{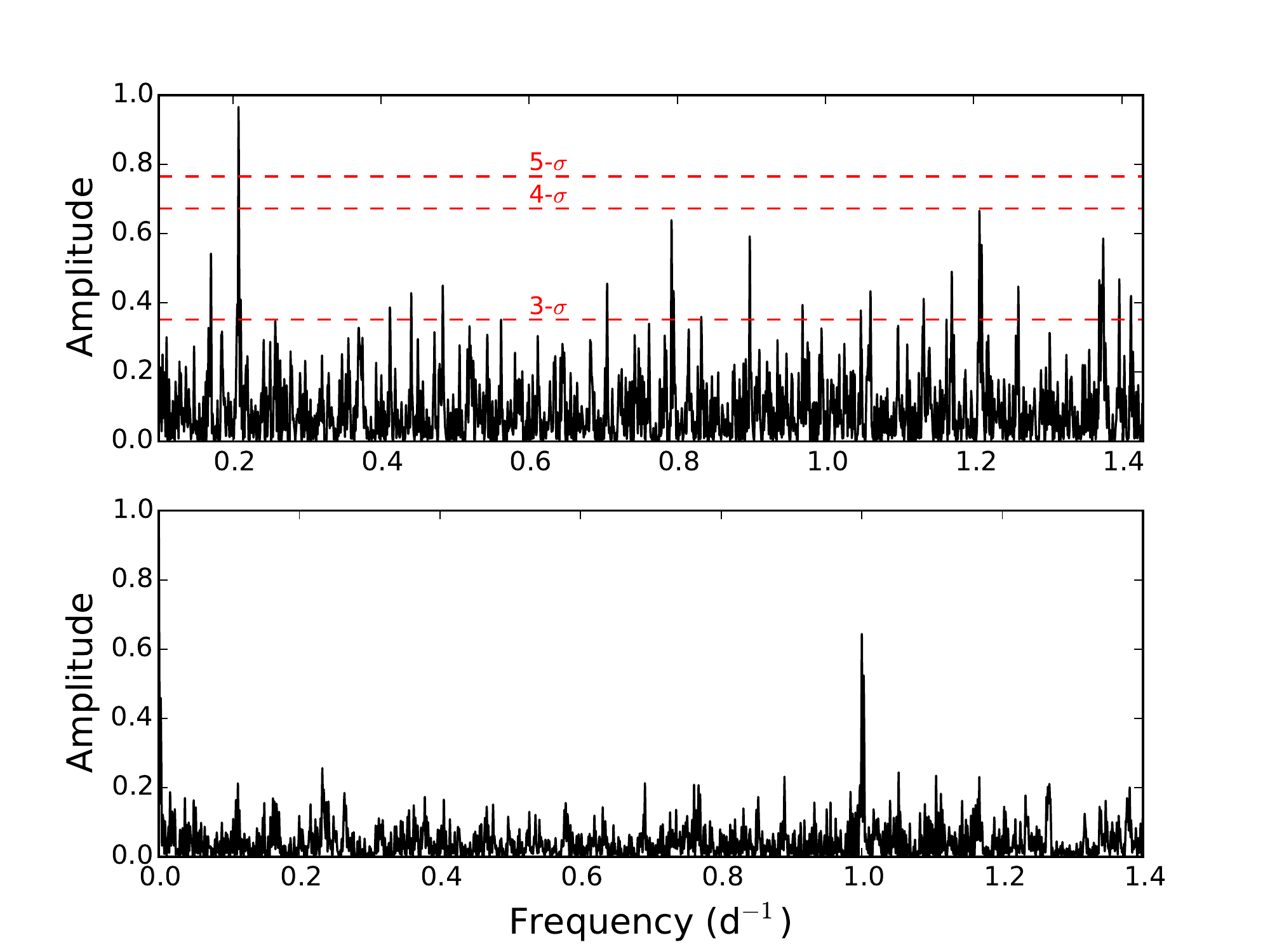}
      \includegraphics[scale=0.45,bb=0 0 576 432]{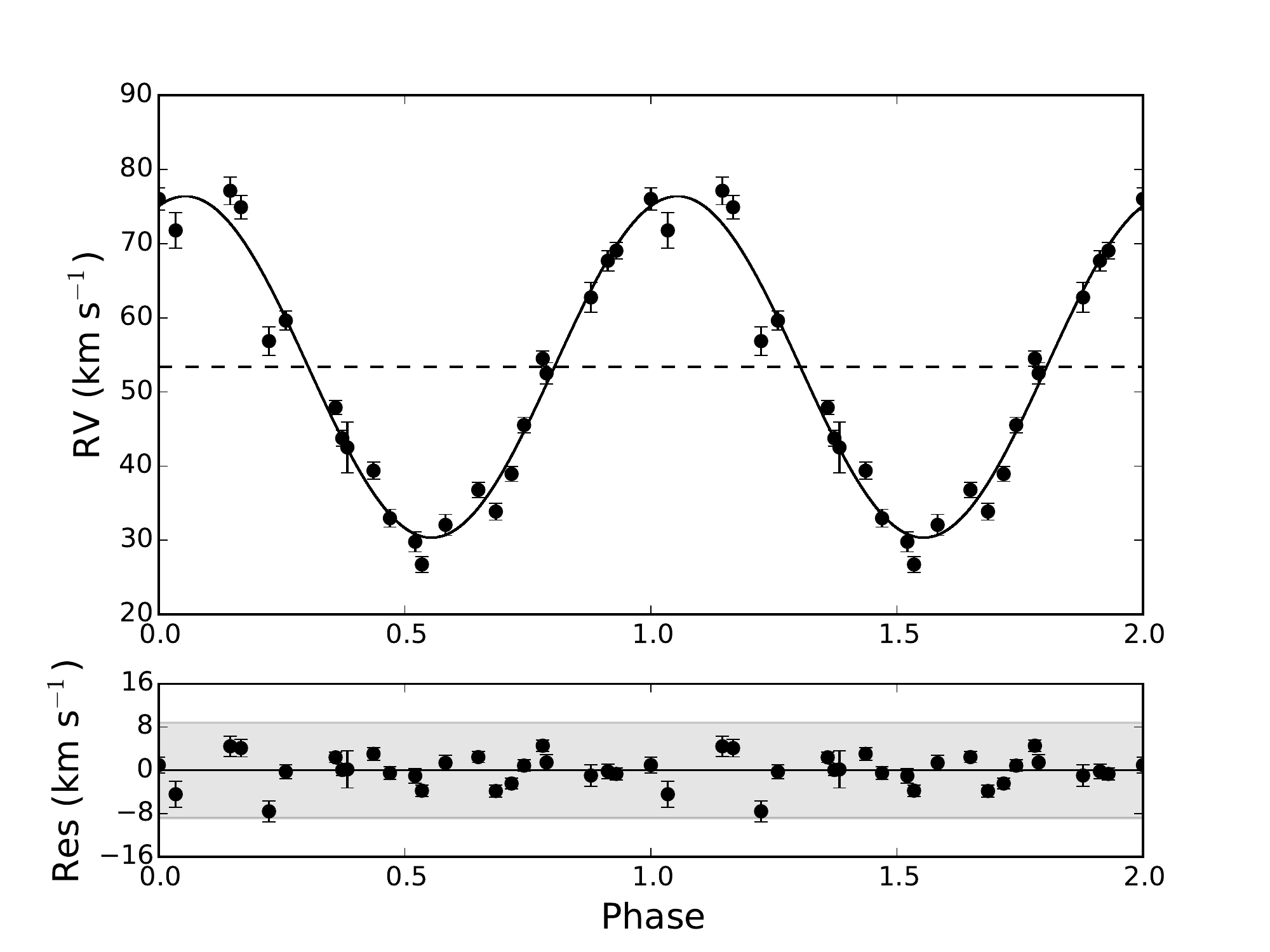}
   \end{center}
   \caption{\emph{(Top panel)} Lomb-scargle periodogram of the SALT HRS HeII $\lambda$4540 RV measurements (top half) and the window function (bottom half). The strongest peak at $f=0.208$ d$^{-1}$ corresponds to the orbital period. \emph{(Bottom panel)} SALT HRS RV measurements phased with the orbital period. The solid line respresents the Keplerian orbit fit and the shaded region indicates the resdiuals are within 3$\sigma$ of the fit where $\sigma=2.94$ km s$^{-1}$.} 
   \label{fig:rv}
\end{figure}

\begin{figure}
   \begin{center}
      \includegraphics[scale=0.55,bb=0 0 401 300]{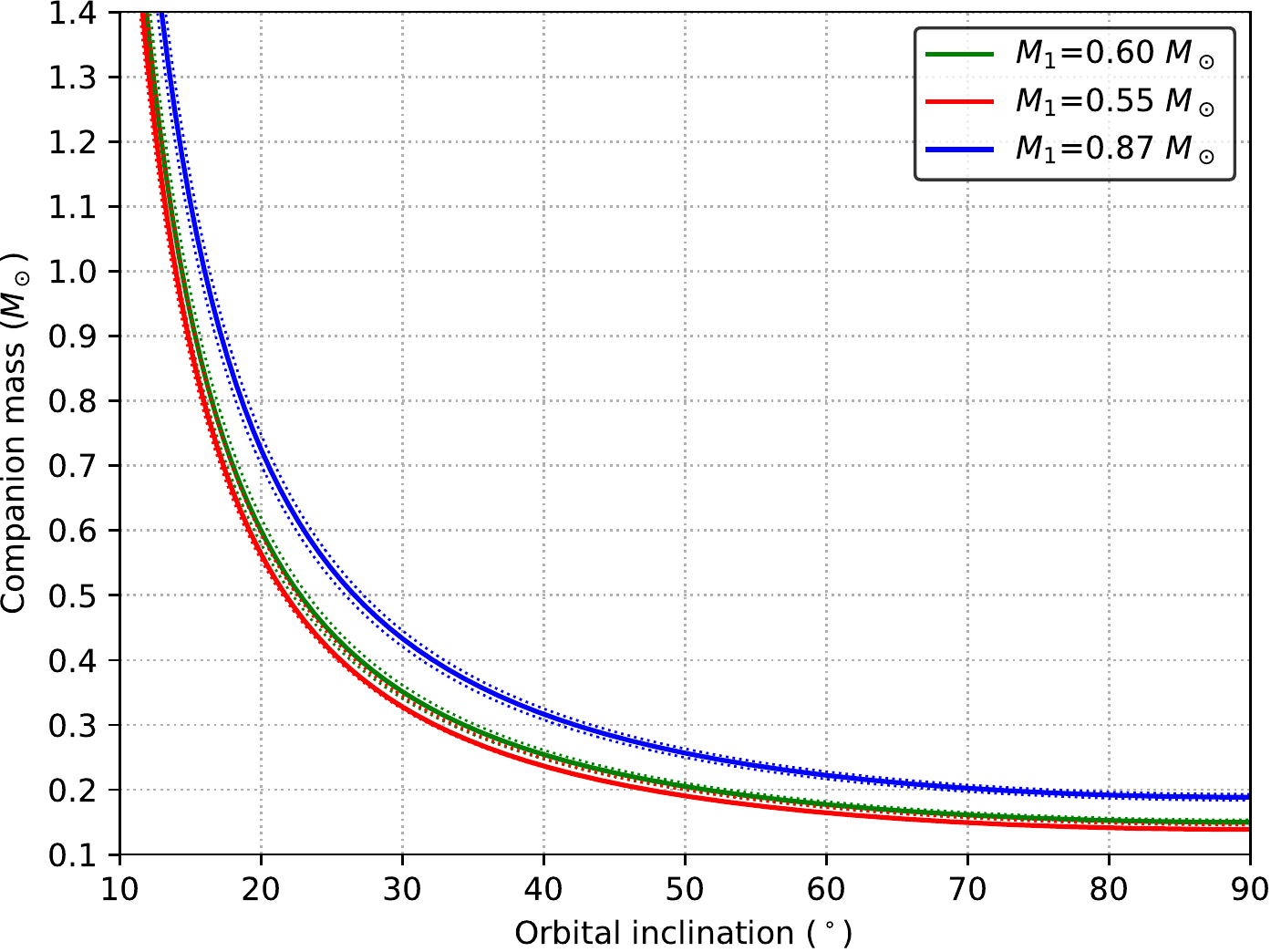}
   \end{center}
   \caption{Companion masses permitted by the mass function in Table \ref{tab:orbit}  for an assumed primary mass of $M_1=0.60^{+0.27}_{-0.05}$ $M_\odot$. The dotted lines indicate the corresponding uncertainty in the mass function.}
   \label{fig:masses}
\end{figure}

The SALT HRS RV measurements were analysed using a Lomb-Scargle periodogram (Press et al. 1992). The strongest peak in the periodogram displayed in Fig. \ref{fig:rv} is at $f=0.208$ d$^{-1}$ and corresponds to an orbital period of 4.81 d. This orbital period was used as the basis for fitting a Keplerian orbit model that was built using a least-squares minimisation method applied to the phase-folded data. Figure \ref{fig:rv} also shows the RV measurements phased with the orbital period, together with the Keplerian orbit fit and the residuals. Table \ref{tab:orbit} lists the orbital parameters determined from Monte Carlo simulations (for details see Miszalski et al. 2018a). An eccentric orbit is not supported by the Lucy \& Sweeney (1971) diagnostic test and we therefore fixed a circular orbit. Assuming the primary mass determined in Sect. \ref{sec:atmos}, Figure \ref{fig:masses} shows possible companion masses permitted by the mass function as a function of the orbital inclination.

A detailed spatiokinematic study of the nebula is required to constrain the orbital inclination of the binary which is expected to match the nebula orientation (Hillwig et al. 2016). However, the apparent nebula morphology (Sect. \ref{sec:morph}) permits a first estimate of the orbital inclination. The bipolar lobes visible in Fig. \ref{fig:fp}d could be produced by a bipolar nebula at an inclination of $\sim$20 deg to the line of sight (e.g. Model A in Figure 2 of Miszalski et al. 2009b). The apparent broken ring feature (Sect. \ref{sec:morph}) may also be interpreted as the waist of a bipolar nebula viewed near pole-on (e.g. Garc{\'{\i}}a-D{\'{\i}}az et al. 2009). If the orbital inclination were $\sim$20 deg, the companion mass in Fig. \ref{fig:masses} would suggest a companion mass of $\sim$0.6 $M_\odot$, corresponding to a WD or a late K-type companion. At greater orbital inclinations the companion mass would correspond to an M-dwarf companion, however we note that this configuration with an 4.8 orbital period would be considered anomalous in the context of the bias-corrected orbital period distribution of WD main-sequence binaries (Nebot G{\'o}mez-Mor{\'a}n et al. 2011; see also Miszalski et al. 2019b). 

\begin{table*}
   \centering
   \caption{Orbital parameters of the binary nucleus of Sp~3 derived from the best-fitting Keplerian orbit to HeII $\lambda$4540 measurements.}
   \label{tab:orbit}
   \begin{tabular}{lc}
      \hline\hline
      Orbital period (d)        & $4.815317\pm0.000664$\\
      Eccentricity $e$ (fixed)       & 0.00 \\
      Radial velocity semi-amplitude $K$ (km s$^{-1}$)  & $22.92\pm$0.51 \\
      Systemic velocity $\gamma$ (km s$^{-1}$) & $52.86\pm$0.36\\
      Epoch at radial velocity minimum $T0$ (d)            & $2457892.840549\pm$0.000664\\
      Root-mean-square residuals of Keplerian fit (km s$^{-1}$) & 2.94 \\
      Separation of primary from centre of mass $a_1\sin i$ (au) & $0.01013\pm$0.00023\\
      Mass function $f(M)$ (M$_\odot$) & $0.00598\pm$0.00040\\
      \hline

   \end{tabular}
\end{table*}

\subsection{Nebular parameters and chemical abundances}
\label{sec:chem}
\begin{figure}
   \begin{center}
      \includegraphics[scale=0.5,bb=0 0 453 336]{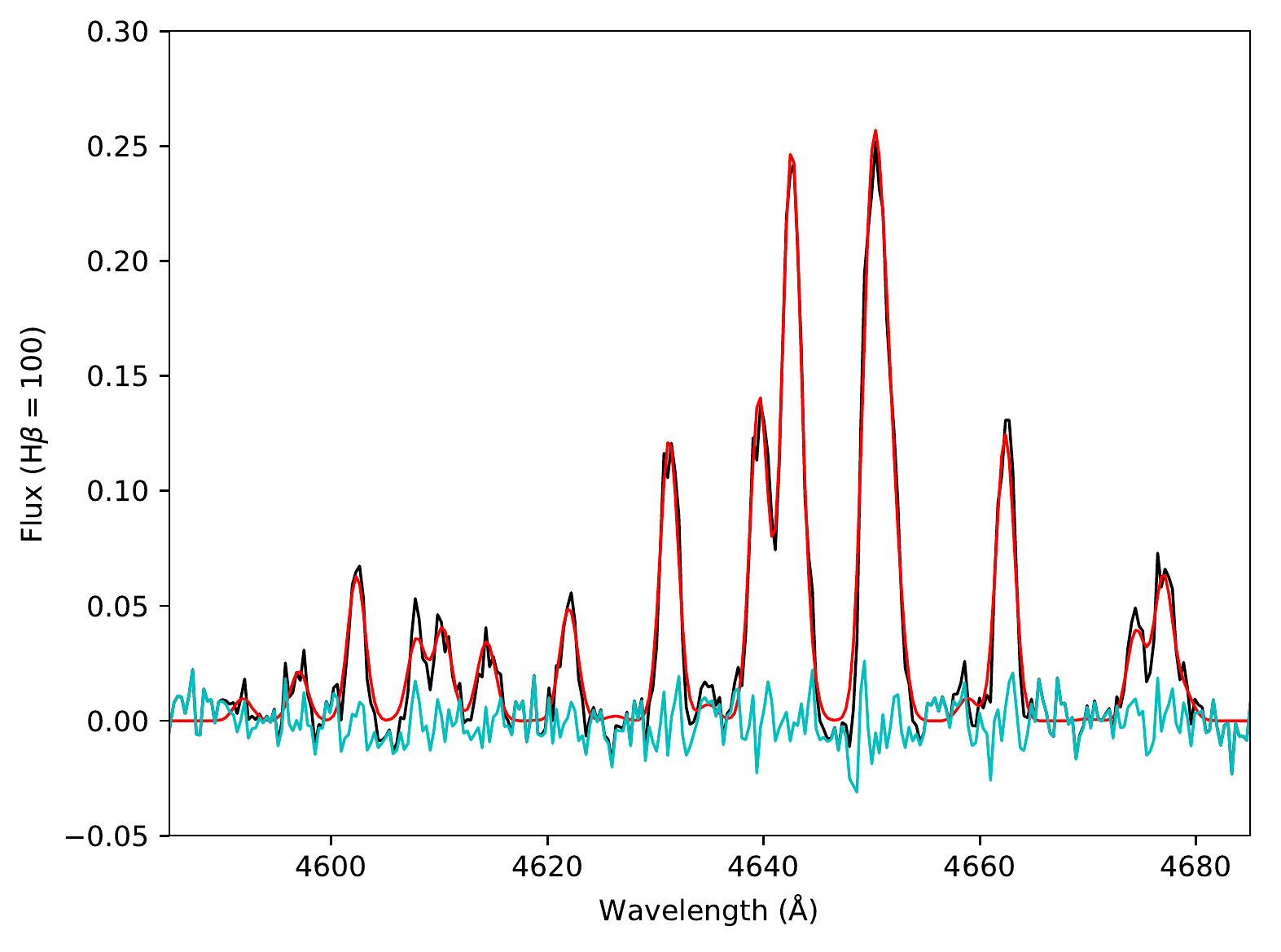}
   \end{center}
   \caption{The continuum-subtracted region of the PG2300 spectrum of Sp~3 near 4650 \AA\ showing the observed spectrum (black), the \textsc{alfa} fit (red) and the residuals (cyan). The intensities have been normalised such that the integrated flux of H$\beta=100$.}
   \label{fig:orlfit}
\end{figure}

We measured emission line fluxes from the RSS longslit spectra using the automated line fitting algorithm program \textsc{alfa} (Wesson 2016). Each spectrum was analysed in two separate halves by \textsc{alfa} to better fit the emission line profiles. This was necessary to account for the slowly varying resolution with wavelength introduced by the volume-phase holographic gratings of RSS. Unsaturated measurements of the H$\alpha$ and [N~II] $\lambda$6548, 6583 \AA\ emission lines were taken from the 180 s PG900 spectrum. Figure \ref{fig:orlfit} shows the \textsc{alfa} fits to the observed region around 4650 \AA\ which contains several nebular recombination lines due to O~II, N~II and C~III. 

We considered two possibilities to join the PG2300 and PG900 spectra into a single representative spectrum for chemical abundance analysis by the nebular empirical analysis tool \textsc{neat} (Wesson et  al. 2012). First, we considered matching the measured fluxes of He~I $\lambda$4471 \AA\ in the overlap region, however this did not result in consistent measurements of the interstellar extinction from the H$\alpha$/H$\beta$ and H$\gamma$/H$\beta$ ratios. The He~I temperatures based on the He~I 5876/4471 and He~I 6678/4471 ratios were also inconsistent with each other and with the O~II temperature. We therefore followed the approach taken by Wesson et al. (2018) where the scale factor was determined with \textsc{neat} such that the H$\alpha$/H$\beta$ and H$\gamma$/H$\beta$ ratios gave consistent measurements of the extinction. A modest scale factor of 0.9685 times the PG2300 spectrum was determined. All lines bluer than 4800 \AA\ in the final joined spectrum were then taken from the PG2300 spectrum scaled by this factor. The joined spectrum was analysed by \textsc{neat} and the identified emission lines are provided in Table \ref{tab:lines1}. The average logarithmic extinction at $H\beta$, $c(H\beta)=0.06^{+0.05}_{-0.04}$, corresponding to $E(B-V)=0.09^{+0.07}_{-0.06}$ (Howarth 1983) is consistent with the previous estimate of $E(B-V)=0.16$ (Ciardullo et al. 1999). Table \ref{tab:diag} presents the electron density and temperature diagnostics, while Tab. \ref{tab:abundances} contains the ionic and total abundances plus calculated O$^{2+}$ and N abundance discrepancy factors (adfs). The adfs are calculated as the ratio of the abundances determined from optical recombination lines (ORLs) to those determined from collisionally excited lines (CELs).

\begin{table}
\caption{Electron density and temperature diagnostics.}
\label{tab:diag}
\centering
\begin{tabular}{ll}
\hline
\hline
Density diagnostic & $n_e$ (cm$^{-3}$) \\ \hline
{}[O~{\sc ii}] 3729/3726 & ${   750}^{+   140}_{  -120}$ \\
{}[Cl~{\sc iii}] 5537/5517 & ${  1040}^{+   730}_{  -610}$ \\
{}[S~{\sc ii}] 6731/6717             & ${   640}^{+   270}_{  -210}$ \\
OII 4649/4089, 4649/4662 & ${   690}^{+   520}_{  -690}$ \\
\hline Temperature diagnostic & $T_e$ (K)\\ \hline

{}[N~{\sc ii}] $(6548+6584)/5754$ & ${  8230}\pm{   160}$ \\
{}[O~{\sc iii}] $(4959+5007)/4363$ & ${7240}\pm{150}$ \\
He I 5876/4471        & ${  2590}^{+  1450}_{  -910}$ \\
He I 6678/4471        & ${  3620}^{+  3940}_{ -1670}$ \\
OII 4649/4089, 4649/4662& ${  3600}^{+  5210}_{ -3600}$ \\
\hline
\hline
\end{tabular}
\end{table}

\begin{table}
\caption{Ionic and total abundances for Sp~3.}
\label{tab:abundances}
\centering
\begin{tabular}{ll}
\hline
\hline
 \multicolumn{2}{l}{CEL abundances}\\ \hline
N$^{+}$/H                           & ${  2.01\times 10^{ -5}}^{+  2.20\times 10^{ -6}}_{ -1.90\times 10^{ -6}}$ \\
icf(N)                              & ${  9.62}^{+  1.53}_{ -1.32}$ \\
N$^{}$/H                            & ${  1.93\times 10^{ -4}}^{+  2.80\times 10^{ -5}}_{ -2.40\times 10^{ -5}}$ \\
O$^{+}$/H                           & ${  5.03\times 10^{ -5}}^{+  5.50\times 10^{ -6}}_{ -5.00\times 10^{ -6}}$ \\
O$^{2+}$/H                          & ${  1.20\times 10^{ -4}}^{+  1.30\times 10^{ -5}}_{ -1.20\times 10^{ -5}}$ \\
icf(O)                              & ${  1.00}\pm{  0.00}$ \\
O$^{}$/H                            & ${  1.71\times 10^{ -4}}^{+  1.40\times 10^{ -5}}_{ -1.30\times 10^{ -5}}$ \\
Ne$^{2+}$/H                         & ${  4.11\times 10^{ -5}}^{+  5.10\times 10^{ -6}}_{ -4.60\times 10^{ -6}}$ \\
icf(Ne)                             & ${  2.73}^{+  0.19}_{ -0.18}$ \\
Ne$^{}$/H                           & ${  1.12\times 10^{ -4}}^{+  1.10\times 10^{ -5}}_{ -1.00\times 10^{ -5}}$ \\
Ar$^{2+}$/H                         & ${  1.24\times 10^{ -6}}^{+  1.20\times 10^{ -7}}_{ -1.10\times 10^{ -7}}$ \\
icf(Ar)                             & ${  1.16}^{+  0.03}_{ -0.03}$ \\
Ar$^{}$/H                           & ${  1.44\times 10^{ -6}}^{+  1.60\times 10^{ -7}}_{ -1.40\times 10^{ -7}}$ \\
S$^{+}$/H                           & ${  5.50\times 10^{ -7}}^{+  5.50\times 10^{ -8}}_{ -5.00\times 10^{ -8}}$ \\
icf(S)                              & ${  6.95}^{+  0.76}_{ -0.68}$ \\
S$^{}$/H                            & ${  3.83\times 10^{ -6}}^{+  4.60\times 10^{ -7}}_{ -4.10\times 10^{ -7}}$ \\
Cl$^{2+}$/H                         & ${  6.84\times 10^{ -8}}^{+  8.30\times 10^{ -9}}_{ -7.40\times 10^{ -9}}$ \\
icf(Cl)                             & ${  1.35}^{+  0.02}_{ -0.02}$ \\
Cl$^{}$/H                           & ${  9.28\times 10^{ -8}}^{+  1.20\times 10^{ -8}}_{ -1.07\times 10^{ -8}}$ \\
\hline
 \multicolumn{2}{l}{ORL abundances}\\ \hline
 He$^{+}$/H                          & ${  1.31\times 10^{-1}}\pm{  5.00\times 10^{ -3}}$ \\
He/H                                & ${  1.31\times 10^{-1}}\pm{  5.00\times 10^{ -3}}$ \\
C$^{2+}$/H                          & ${  2.17\times 10^{ -3}}\pm{  5.00\times 10^{ -5}}$ \\
C$^{3+}$/H                          & ${  6.40\times 10^{ -5}}\pm{  1.41\times 10^{ -5}}$ \\
icf(C)                              & ${  1.16}^{+  0.02}_{ -0.01}$ \\
C/H                                 & ${  2.59\times 10^{ -3}}^{+  8.00\times 10^{ -5}}_{ -7.00\times 10^{ -5}}$ \\
N$^{2+}$/H                          & ${  1.83\times 10^{ -3}}\pm{  7.00\times 10^{ -5}}$ \\
icf(N)                              & ${  1.00}\pm{  0.00}$ \\
N/H                                 & ${  1.83\times 10^{ -3}}\pm{  7.00\times 10^{ -5}}$ \\
O$^{2+}$/H                          & ${  2.95\times 10^{ -3}}^{+  3.70\times 10^{ -4}}_{ -2.90\times 10^{ -4}}$ \\
icf(O)                              & ${  1.42}^{+  0.06}_{ -0.06}$ \\
O/H                                 & ${  4.19\times 10^{ -3}}^{+  5.70\times 10^{ -4}}_{ -4.40\times 10^{ -4}}$ \\
 \hline
 \multicolumn{2}{l}{Abundance discrepancy factors}\\ \hline
adf (O$^{2+}$/H)                    & ${ 24.6}^{+  4.1}_{ -3.4}$ \\
adf (N/H)                           & ${  9.5}^{+  1.4}_{ -1.2}$ \\
\hline
\hline
\end{tabular}
\end{table}

\section{DISCUSSION}
\label{sec:discussion}
\subsection{Distance and likelihood of visual companion physical association}
\label{sec:distance}
Could the visual companion identified by Ciardullo et al. (1999) be physically associated with the newly discovered post-CE central star, therefore making the nucleus a triple system? Ciardullo et al. (1999) argued that the small separation (0.31\arcsec) and characteristics of the companion are in agreement with the prior statistical distance estimates of the nebula, suggesting a true physical association. Stanghellini \& Haywood (2010) estimated a distance of 1.92$\pm$0.38 kpc based on the nebula properties. Frew et al. (2016) determined a G0V spectral type and a spectroscopic distance of $2.22^{+0.61}_{-0.48}$ kpc for the visual companion. Frew et al. (2016) also estimated a distance based on the nebular properties of 2.11$\pm$0.60 kpc, where the spectroscopic distance to the visual companion was used as a basis for including Sp~3 as a calibrator for their distance estimation method. 

Despite the apparent agreement between all these distances, it is worthwhile to consider other independent distance measurements to check the suspected physical association of the visual companion, especially given the difficulties associated with estimating PN distances (Frew et al. 2016). Here we consider distances estimated from the recent \emph{Gaia} DR2 parallax measurement of the central star (Gaia Collaboration et al. 2018a) and the photospheric parameters of the primary we have derived from our TMAP NLTE analysis (Sect. \ref{sec:atmos}).

Table \ref{tab:gaia} collates parameters recorded for the nucleus of Sp~3 in the second data release (DR2, Gaia Collaboration et al. 2018a) of the \emph{Gaia} mission (Gaia Collaboration et al. 2016) and other derived quantities.\footnote{A separate detection of the visual companion was not recorded in the \emph{Gaia} DR2 catalogue.} The \emph{Gaia} DR2 astrometry is affected by many systematic effects as discussed in papers associated with the data release (e.g. Lindegren et al. 2018; Gaia Collaboration et al. 2018b; Arenou et al. 2018). Distance determination is not necessarily a straight-forward exercise of taking the reciprocal of the parallax (Luri et al. 2018) and a Bayesian approach is the preferred method (Bailer-Jones et al. 2018). Furthermore, in the case of PNe the nebula may introduce additional biases (Kimeswenger \& Barr\'ia 2018), though the full extent of such biases is yet to be determined. The catalogue of Bailer-Jones et al. (2018) provides robust distance estimates for \emph{Gaia} DR2 sources. Distance estimates in the catalogue include lower and upper boundaries of the highest density interval around the mode of the posterior with probability $p=0.6827$. A Gaussian posterior would correspond to an uncertainty of $\pm1\sigma$ in the distance. 

\begin{table*}
   \centering
   \caption{\emph{Gaia} DR2 parameters and derived quantities for the central star of Sp~3. The parallax $\varpi$ includes a zero-point correction of $+0.029$ mas (see Lindegren et al. 2018). Filters adopted by Gaia Collaboration et al. (2018b), indicated by inequalities and thresholds (enclosed in parentheses) for the relevant values, are all satisifed in the cases shown here.}
   \begin{tabular}{ll}
      \hline
      \verb|source_id| & 6702910370854823296 \\
      $\varpi$ (mas) & $-0.431$ \\
      $\sigma_\varpi$ (mas) & $0.109$\\
      $\sigma_\varpi/\varpi$ & $-0.253$ \\
      $1/\varpi=d$ (kpc) & 2.32$^{+0.79}_{-0.47}$ \\ 
      $G$ (mag) & 13.0901$\pm$0.0012 \\
      $G_\mathrm{BP}$ (mag) & 12.8947$\pm$0.0107  \\
      $G_\mathrm{RP}$ (mag) & 13.2083$\pm$0.0022  \\
      $G_\mathrm{BP}-G_\mathrm{RP}$ (mag) & $-0.31$ \\
      \verb|astrometric_n_good_obs_al| $=\nu'$ & 143  \\
      \verb|visibility_periods_used| & 11 ($>8$)\\
      \verb|astrometric_chi2_al| $=\chi^2$ & 2424.68  \\
      \verb|astrometric_excess_noise| (mas) & 0.68 ($<1.0$) \\
      \verb|phot_g_mean_flux_over_error| & 915.73 ($>50$)  \\
      \verb|phot_bp_mean_flux_over_error| & 101.73 ($>20$)\\
      \verb|phot_rp_mean_flux_over_error| & 492.83 ($>20$) \\
      \verb|phot_bp_rp_excess_factor| $=E$ & 1.26  \\
      $\sqrt{\chi^2/(\nu'-5)}$ $<$ $1.2\,\mathrm{max}\left(1,\exp(-0.2\,(G-19.5))\right)$ & 4.19 $<$ 4.32  \\
      $1.0+0.015\left(G_\mathrm{BP}-G_\mathrm{RP}\right)^2$ $<$ $E$& 1.00 $<$ 1.26  \\
      $E$ $<$ $1.3+0.06\left(G_\mathrm{BP}-G_\mathrm{RP}\right)^2$ & 1.26 $<$ 1.31  \\
      \hline
   \end{tabular}
   \label{tab:gaia}
\end{table*}

In the case of Sp~3, the Bailer-Jones et al. (2018) estimate of $r_\mathrm{est}=11.2$ kpc, with boundaries of $r_\mathrm{lo}=8.1$ kpc and $r_\mathrm{hi}=15.4$ kpc, unfortunately appears to be too distant. At 11.2 kpc the 34\arcsec\ nebula radius (Sect. \ref{sec:morph}) would correspond to $\sim$1.85 pc, considerably larger than most PNe (Frew et al. 2016). The morphology of such a large nebula would more closely resemble evolved, low surface-brightness PNe, e.g. PFP1 (Pierce et al. 2004), inconsistent with the observed appearance of Sp~3 (Sect. \ref{sec:morph}). Given the implausible nature of this result, we have no other recourse but to estimate the distance as the reciprocal of the parallax to obtain $d=2.32^{+0.79}_{-0.47}$ kpc. Despite the difficulties associated with this approach (Luri et al. 2018), we are somewhat reassured by the fact that the parameters in Tab. \ref{tab:gaia} satisfy several quality criteria filters, namely in the form of inequalities and thresholds, that are applied to \emph{Gaia} DR2 data of large samples before analysis (Sect. 2.1 of Gaia Collaboration et al. 2018b; see also Lindegren et al. 2018 and Arenou et al. 2018). 

\begin{figure}
   \includegraphics[scale=0.95,bb=20 20 300 253]{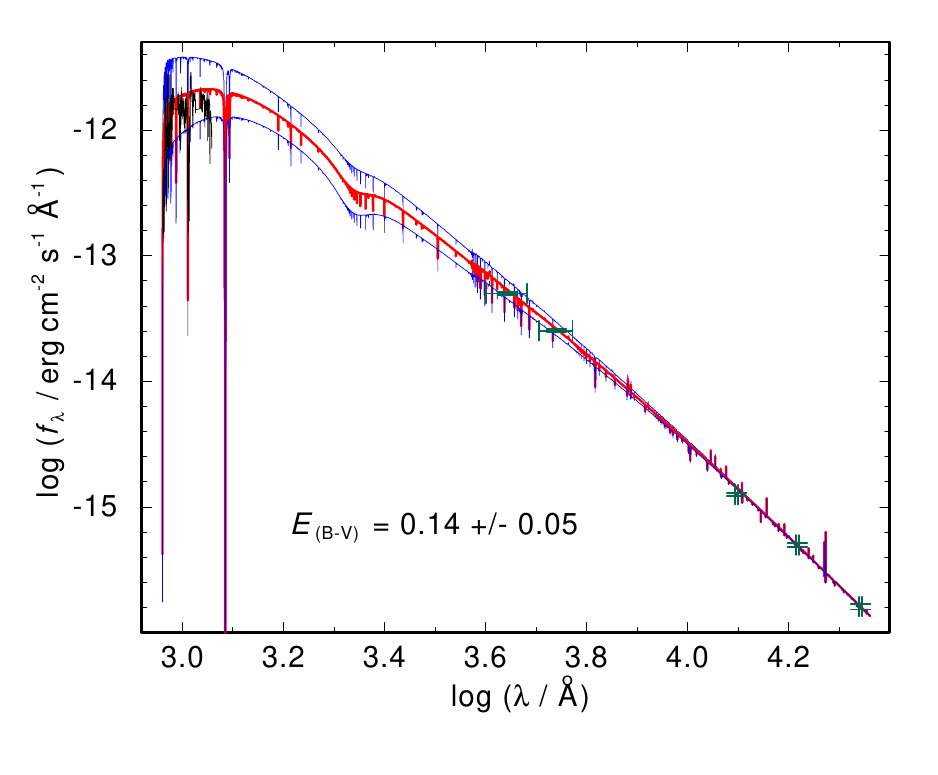}
   \caption{Determination of \ebv\, for the CSPN of Sp~3 using the \emph{FUSE} spectrum (Id B032080100000 retrieved from the MAST archive; black line) and the $B$ and $V$ (Zacharias et al. 2013; Henden et al. 2016) and the 2MASS $J$, $H$, and $K_s$ magnitudes (Cutri et al. 2003). The model has \Teffw{68\,000} and \loggw{4.6} and is normalized to the $K_s$ magnitude (red line). The blue lines indicate the \ebv\ error range.}
   \label{fig:ebv}
\end{figure}

We have also calculated the spectroscopic distance of the CSPN of Sp~3 using the flux calibration of Heber et al. (1984) for $\lambda_\mathrm{eff} = 5454$ \AA, 
$$d[\mathrm{pc}]=7.11 \times 10^{-4} \cdot \sqrt{H_\nu\cdot M \times 10^{0.4\, m_{\mathrm{v}_0}-\log g}} \,\, ,$$
\noindent
with $m_\mathrm{V_o} = m_\mathrm{V} - 2.175 c$, $c = 1.47 E_\mathrm{B-V}$, and 
the Eddington flux $H_\nu$ ($1.52\times 10^{-3}$ erg/cm$^{2}$/s/Hz) at 5454\,\AA\,\,of our final model atmosphere.
We use $m_{\mathrm{V}}=12.89$ that was measured by Zacharias et al. (2013) and Henden et al. (2016).
With $E_\mathrm{B-V}=0.14 \pm 0.05$ (Fig.\,\ref{fig:ebv}) and $M = 0.60^{+0.27}_{-0.05}$\,\Msol,
we derived 
$d=2.8^{+0.8}_{-0.7}$\,kpc. Regarding the He~II $\lambda$4686.06 \AA\, discrepancy (see Sect. \ref{sec:atmos}), we find better agreement between model and observation at about \Teffw{80\,000}. However, then the star is already located very close to the Eddington limit (Fig.\,\ref{fig:evolution}) and, thus, would be more massive $M = 0.83^{+0.18}_{-0.08}$\,\Msol\ and at a much further distance of $d=4.0^{+0.9}_{-1.2}$\,kpc. This distance is around two times further than the other distance estimates and seems unlikely. 

Table \ref{tab:dist} provides a summary of the various distance estimates to Sp~3. While the actual veracity of the distance obtained from the reciprocal of the parallax may only become clear once additional observations and improved data processing are available from future \emph{Gaia} data releases, the overall picture is one that clearly supports a likely physical association between the visual companion and the post-CE binary nucleus of Sp~3.

\begin{table*}
   \centering
   \caption{A summary of various distances to Sp~3.}
   \label{tab:dist}
   \begin{tabular}{lll}
      \hline
      Quantity & Distance (kpc) & Reference  \\
      \hline
      $d_\mathrm{nebula}$  & 1.92$\pm$0.38 & Stanghellini \& Haywood (2010)\\
      $d_\mathrm{spec,tertiary}$ & $2.22^{+0.61}_{-0.48}$ & Frew et al. (2016)\\
      $d_\mathrm{nebula}$  & 2.11$\pm$0.60& Frew et al. (2016)\\
      $r_\mathrm{est}$ & 11.2$^{+4.2}_{-3.1}$ & Bailer-Jones et al. (2018)\\
      $1/\varpi$ & 2.32$^{+0.79}_{-0.47}$ & Gaia Collaboration et al. (2018a); This work \\ 
      $d_\mathrm{gravity}$ & 2.8$^{+0.8}_{-0.7}$ & This work\\
      \hline
   \end{tabular}
\end{table*}

\subsection{Chemical abundances}
The most prominent result is that the adf(O$^{2+}$) of $24.6^{+4.1}_{-3.4}$ lies in the `extreme' range (adf $>$ 10) for PNe (Wesson et al. 2018). Wesson et al. (2018) identified several trends with adf(O$^{2+}$) that post-CE PNe follow concerning the [S~II] and [O~II] electron densities, as well as the O/H and N/H abundances. The location of Sp~3 with its low nebular densities and `extreme' adf is evidently consistent with these trends, though their cause is not yet clear (Wesson et al. 2018). 

\begin{figure}
   \begin{center}
      \includegraphics[scale=0.55,bb=0 0 418.02 318.590125]{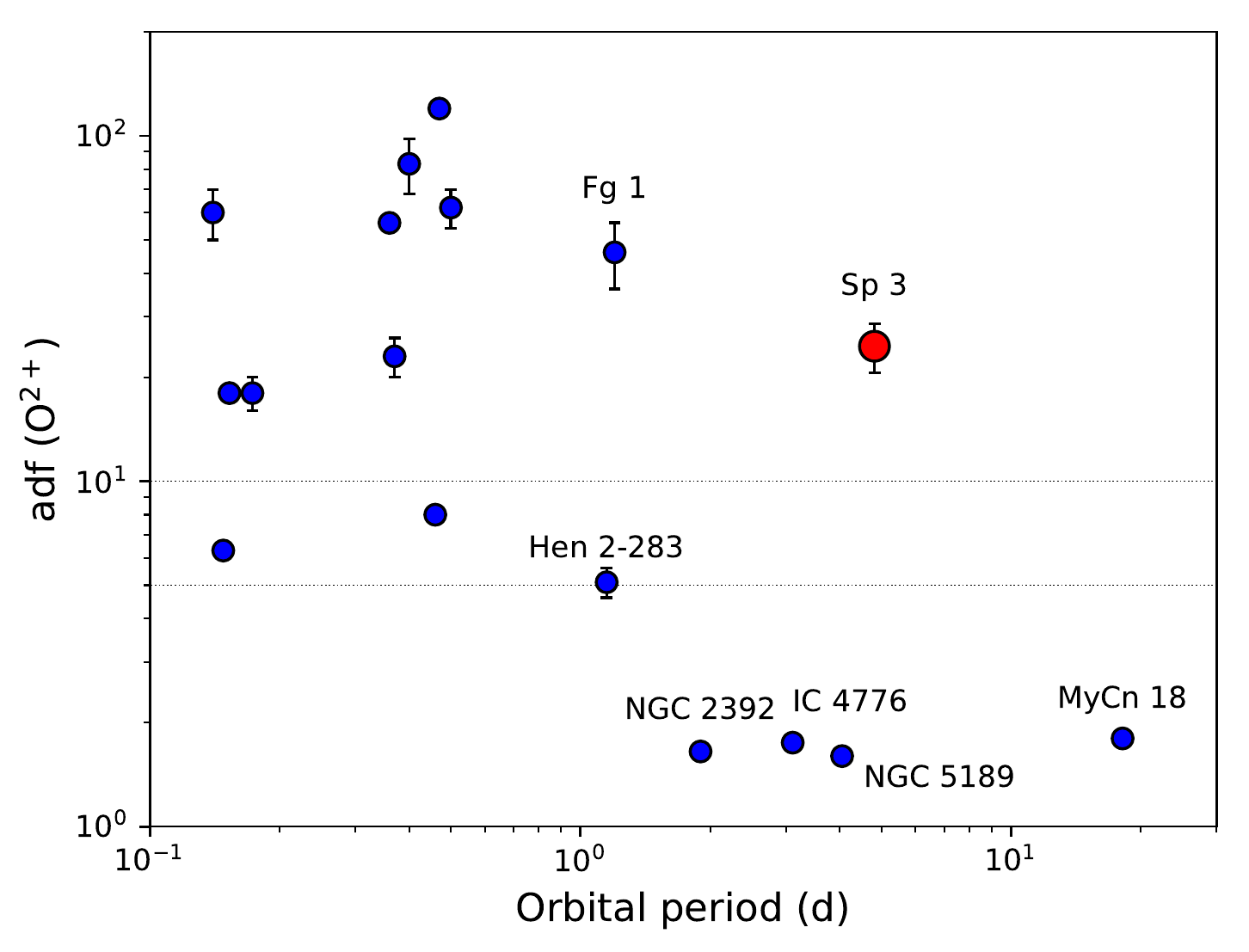}
   \end{center}
   \caption{The location of Sp~3 amongst other post-CE PNe with measured adfs and orbital periods. Post-CE PNe with orbital periods in excess of 1.0 d are labelled. The dotted lines mark the thresholds of Wesson et al. (2018) indicative of `normal' (adf $<$ 5), `elevated' ($5<$ adf $<10$) and `extreme' (adf $>10$) adfs. Sp~3 occupies a previously unpopulated part of the parameter space.}
   \label{fig:adfs}
\end{figure}

Wesson et al. (2018) also found that only post-CE PNe with orbital periods less than $\sim$1.15 d demonstrated `extreme' adfs. Figure \ref{fig:adfs} depicts the adf(O$^{2+}$) as a function of orbital period constructed using data from Tab. 6 of Wesson et al. (2018). We have added Sp~3, together with MyCn~18 ($P=18.15$ d, Miszalski et al. 2018b; adf$(\mathrm{O}^{2+})=1.8$, Tsamis et al. 2004) and NGC~2392 ($P=1.9$ d, Miszalski et al. 2019a; adf$(\mathrm{O}^{2+})=1.65$, Zhang et al. 2012). The orbital period of IC~4776 was revised down to 3.11 d (Miszalski et al. 2019b) and we excluded Hen~2-161 whose orbital period is uncertain. The 4.8 d orbital period of Sp~3 clearly breaches the expected tendency for multiple day orbital period post-CE to show normal adfs (Wesson et al. 2018), making it a clear outlier in Fig. \ref{fig:adfs}. 

The extreme adf of Sp~3 emphasises the presence of strong selection effects in the known post-CE PN orbital period distribution. We consider any relationships inferred between the adf and orbital period to therefore not be meaningful, especially given the still very small population of post-CE PNe with determined adfs. These selection effects are primarily determined by the use of photometric monitoring to discover most post-CE PNe (e.g. Miszalski et al. 2009a). Indeed, we note that all post-CE PNe with orbital periods above 1.0 d in Fig. \ref{fig:adfs} were identified via RV monitoring except Hen~2-283!

The He abundance ($12+\log(\mathrm{He/H})=11.11$ dex) and $\log(\mathrm{N/O})=0.05$ dex are typical of Type I PNe (Kingsburgh \& Barlow 1994) that are believed to form from more massive progenitors ($M\sim3 M_\odot$, Karakas \& Lattanzio 2014), making Sp~3 one of very few Type I PNe amongst post-CE PNe (Corradi et al. 2014). The apparent bipolar morphology (Sect. \ref{sec:morph}) is also consistent with the Type I abundance pattern (Corradi \& Schwarz 1995). The oxygen abundance ($0.46$ dex below Solar, Asplund et al. 2009) and the height below the Galactic plane ($z=-0.57$ kpc assuming $d=2.32$ kpc, Sect. \ref{sec:distance}) both suggest Sp~3 belongs to the thick disk of the Galaxy (e.g. Robin et al. 2014). 

\section{CONCLUSIONS}
\label{sec:conclusion}
We have presented a SALT study of the PN Sp~3 and its central star for which Ciardullo et al. (1999) previously identifed to have a visual companion located 0.31\arcsec\ away. Radial velocity measurements obtained with SALT HRS reveal the central star to be a post-CE binary with an orbital period of 4.81 d. The spectroscopic distance of the visual companion ($2.22^{+0.61}_{-0.48}$ kpc, Frew et al. 2016) is in agreement with estimates of the distance to Sp~3 based on the nebula properties ($1.92\pm0.38$ kpc, Stanghellini \& Haywood 2010; $2.11\pm0.60$ kpc, Frew et al. 2016), the \emph{GAIA} DR2 parallax of the central star ($2.32^{+0.79}_{-0.47}$ kpc, Gaia Collaboration et al. 2018a) and the photospheric properties of the central star ($2.8^{+0.8}_{-0.7}$ kpc). This strongly suggests that the visual companion is associated with the post-CE binary nucleus, indicating that nucleus of Sp~3 is a likely triple system. This is the strongest candidate for a triple nucleus of a PN besides the only proven case of NGC~246 (Adam \& Mugrauer 2014). 

Our main conclusions are as follows:

\begin{itemize}
   \item A total of 23 SALT HRS RV measurements find the nucleus of Sp~3 to be a spectroscopic binary with an orbital period of 4.81 d and a RV semi-amplitude of $22.92\pm0.51$ km s$^{-1}$. This is one of the longest orbital periods known in PNe (Miszalski et al. 2019b) and higher than expected for post-CE WD main-sequence binaries (Nebot G{\'o}mez-Mor{\'a}n et al. 2011), further supporting the possibility that there may be a larger population of longer orbital period binary central stars waiting to be found. Sp~3 is the third binary we have identified in the list of RV variables identified by Af{\v s}ar \& Bond(2005), after NGC~1360 (Miszalski et al. 2018a) and NGC~2392 (Miszalski et al. 2019a).
   \item The TMAP NLTE model atmosphere analysis of the SALT HRS spectra show the primary to be a relatively fast rotator ($v_\mathrm{rot}=80\pm20$ km s$^{-1}$) with $T_\mathrm{eff}=68\,000^{+12\,000}_{-6\,000}$ K and $\log g=4.6\pm0.2$. Interpolation with the H-rich stellar evolutionary tracks of Miller Bertolami et al. (2016) show that the central star is relatively close to the Eddington limit with $M=0.60^{+0.27}_{-0.05}$ $M_\odot$ and log ($L/L_\odot$)=$3.85^{+0.55}_{-0.35}$. High-resolution UV spectroscopy is required to further investigate the wind properties identified by previous studies (Gauba et al. 2001; Guerrero \& De Marco 2013) and refine the photospheric parameters.
    \item SALT RSS Fabry-P\'erot H$\alpha$ and [O~III] images are presented of the peculiar nebula, revealing new structures that include faint bipolar lobes, jets and a broken ring that may be the waist of a bipolar nebula. The orientation of the nebula is estimated to be $\sim$20 deg, however detailed a spatiokinematic study is required to properly constrain the orientation. Assuming the orientation matches the orbital inclination (Hillwig et al. 2016) and adopting $M_1=0.60^{+0.27}_{-0.05}$ $M_\odot$, the mass function of the binary central star gives a companion mass of $\sim$0.6 $M_\odot$, corresponding to a WD or possibly K-dwarf companion.
    \item SALT RSS longslit spectroscopy of the nebula was used to determine the nebular chemical abundances. Most surprising is the extreme adf(O$^{2+}$) of $24.6^{+4.1}_{-3.4}$, which does not fit the expected low adf of post-CE PNe with orbital periods long than $\sim$1 day (Wesson et al. 2018). Selection effects are therefore playing a dominant role in the current search for trends amongst post-CE PNe. 
    \item The chemical abundance pattern of the nebula is typical of Type-I PNe ($12+\log(\mathrm{He/H})=11.11$ dex; $\log(\mathrm{N/O})=0.05$ dex), thought to evolve from more massive progenitors (Corradi \& Schwarz 1995; Karakas \& Lattanzio 2014). However, the sub-Solar oxygen abundance and large height below the Galactic plane suggest a thick disk membership for Sp~3. This paradox may be attributable to the still poorly understood influence of rotation ($v_\mathrm{rot}=80\pm20$ km s$^{-1}$) and binarity on AGB nucleosynthesis (e.g. Stasi{\'n}ska et al. 2010; Miszalski et al. 2012; Karakas \& Lattanzio 2014 and ref. therein), though we note that Type-I PNe remain under-represented amongst post-CE PNe (Corradi et al. 2014).
    \item It is unclear whether the triple nature of the nucleus has influenced the nebula morphology given its large separation from the binary component ($\sim$740 au, Ciardullo et al. 1999). The current orbit of the post-CE nucleus is circular, however it is interesting to conjecture that the tertiary component may have induced an eccentric orbit in the past via the Kozai-Lidov mechanism (e.g. Toonen et al. 2016). Further modelling of the potential influence of the triple system on the nebula would be of interest. 
\end{itemize}

\begin{acknowledgements}
   BM acknowledges support from the National Research Foundation (NRF) of South Africa and thanks the Institute of Astronomy at KU Leuven for their hospitality. RM acknowledges support from the Claude Leon Foundation. We thank an anonymous referee for constructive comments that helped improve this paper. This paper is based on observations made with the Southern African Large Telescope (SALT) under programmes 2012-1-RSA\_OTH-010, 2016-2-SCI-034 and 2017-1-MLT-010 (PI: B. Miszalski). We are grateful to our SALT colleagues for maintaining the telescope facilities and conducting the observations. We thank S. Mohamed and D. Kamath for discussions, R. Wesson for assistance with his \textsc{alfa} and \textsc{neat} software, A. Y. Kniazev for making available his HRS pipeline data products, and T.E. Pickering for RSS Fabry-P\'erot technical assistance. Polish participation in SALT is funded by grant No. MNiSW DIR/WK/2016/07. This research has been partly founded by the National Science Centre, Poland, through grant OPUS 2017/27/B/ST9/01940 to JM. KI has been financed by the Polish Ministry of Science and Higher Education Diamond Grant Programme via grant 0136/DIA/2014/43 and by the Foundation for Polish Science (FNP) within the START program. HVW and RM acknowledge support from the Belgian Science Policy Office under contract BR/143/A2/STARLAB. HVW acknowledges additional support from the Research Council of K.U. Leuven under contract C14/17/082. Some of the data presented in this paper were obtained from the Barbara A\@. Mikulski Archive for Space Telescopes (MAST). STScI is operated by the Association of Universities for Research in Astronomy, Inc., under NASA contract NAS5-26555. Support for MAST for non-HST data is provided by the NASA Office of Space Science via grant NNX09AF08G and by other grants and contracts. This research has made use of NASA's Astrophysics Data System and the SIMBAD database, operated at CDS, Strasbourg, France. This research made use of Matplotlib (Hunter 2007) and Astropy,\footnote{http://www.astropy.org} a community-developed core Python package for Astronomy Collaboration et al. (2013, 2018). IRAF is distributed by the National Optical Astronomy Observatory, which is operated by the Association of Universities for Research in Astronomy (AURA) under a cooperative agreement with the National Science Foundation. This work has made use of data from the European Space Agency (ESA) mission \emph{Gaia} (https://www.cosmos.esa.int/gaia), processed by the \emph{Gaia} Data Processing and Analysis Consortium (DPAC, https://www.cosmos.esa.int/web/gaia/dpac/consortium). Funding for the DPAC has been provided by national institutions, in particular the institutions participating in the \emph{Gaia} Multilateral Agreement.

\end{acknowledgements}

\begin{appendix}
   \label{sec:appendix}

 \begin{table*}
    \centering
    \caption{Observed $F(\lambda)$ and dereddened $I(\lambda)$ emission line fluxes for Sp~3.}
 \label{tab:lines1}
    \begin{tabular}{lrlrllllllll}
 \hline
 $ \lambda $ & Ion & $F \left( \lambda \right) $ && $I \left( \lambda \right) $ && Ion & Multiplet & Lower term & Upper term & g$_1$ & g$_2$  \\
 \hline
  3697.75 &   3697.15 &    1.487& $\pm$   0.133&   1.389& $^{  +0.137}_{  -0.152}$ & H~{\sc i}        & H17        & 2p+ 2P*    & 17d+ 2D    &          8 &        *    \\
  3704.46 &   3703.86 &    1.849& $\pm$   0.102&   1.916& $^{  +0.114}_{  -0.122}$ & H~{\sc i}        & H16        & 2p+ 2P*    & 16d+ 2D    &          8 &        *    \\
  3705.62 &   3705.02 &    0.880& $\pm$   0.101&   0.869& $^{  +0.103}_{  -0.117}$ & He~{\sc i}       & V25        & 2p 3P*     & 7d 3D      &          9 &       15    \\
  3712.57 &   3711.97 &    2.272& $\pm$   0.112&   2.375& $^{  +0.129}_{  -0.136}$ & H~{\sc i}        & H15        & 2p+ 2P*    & 15d+ 2D    &          8 &        *    \\
  3722.23 &   3721.63 &    2.608& $\pm$   0.147&   2.696& $^{  +0.059}_{  -0.088}$ & [S~{\sc iii}]    & F2         & 3p2 3P     & 3p2 1S     &          3 &        1    \\
        * &   3721.94 &  *     &             & *     &         & H~{\sc i}        & H14        & 2p+ 2P*    & 14d+ 2D    &          8 &        *    \\
  3726.64 &   3726.03 &   25.991& $\pm$   0.898&  26.200& $^{  +1.100}_{  -1.200}$ & [O~{\sc ii}]     & F1         & 2p3 4S*    & 2p3 2D*    &          4 &        4    \\
  3729.43 &   3728.82 &   21.818& $\pm$   0.960&  22.600& $^{  +1.100}_{  -1.200}$ & [O~{\sc ii}]     & F1         & 2p3 4S*    & 2p3 2D*    &          4 &        6    \\
  3734.98 &   3734.37 &    2.861& $\pm$   0.202&   3.123& $^{  +0.217}_{  -0.233}$ & H~{\sc i}        & H13        & 2p+ 2P*    & 13d+ 2D    &          8 &        *    \\
  3750.76 &   3750.15 &    3.475& $\pm$   0.098&   3.488& $^{  +0.133}_{  -0.138}$ & H~{\sc i}        & H12        & 2p+ 2P*    & 12d+ 2D    &          8 &        *    \\
  3771.24 &   3770.63 &    4.052& $\pm$   0.158&   4.108& $^{  +0.190}_{  -0.199}$ & H~{\sc i}        & H11        & 2p+ 2P*    & 11d+ 2D    &          8 &        *    \\
  3798.52 &   3797.90 &    5.283& $\pm$   0.160&   5.716& $^{  +0.214}_{  -0.223}$ & H~{\sc i}        & H10        & 2p+ 2P*    & 10d+ 2D    &          8 &        *    \\
  3820.24 &   3819.62 &    1.827& $\pm$   0.071&   1.925& $^{  +0.085}_{  -0.089}$ & He~{\sc i}       & V22        & 2p 3P*     & 6d 3D      &          9 &       15    \\
  3835.59 &   3834.89 &    8.279& $\pm$   0.217&   8.535& $^{  +0.173}_{  -0.255}$                                                                                      \\
        * &   3835.39 &  *     &             & *     &         & H~{\sc i}        & H9         & 2p+ 2P*    & 9d+ 2D     &          8 &        *    \\
  3856.73 &   3856.02 &    0.136& $\pm$   0.037&   0.141& $^{  +0.003}_{  -0.004}$ & S~{\sc iii}      & V12        & 3p2 2D     & 4p 2P*     &          6 &        4    \\
        * &   3856.13 &  *     &             & *     &         & O~{\sc ii}       & V12        & 3p 4D*     & 3d 4D      &          4 &        2    \\
  3869.46 &   3868.75 &    7.955& $\pm$   0.243&   8.053& $^{  +0.309}_{  -0.322}$ & [Ne~{\sc iii}]   & F1         & 2p4 3P     & 2p4 1D     &          5 &        5    \\
  3889.36 &   3888.65 &   25.902& $\pm$   0.481&  26.666& $^{  +0.517}_{  -0.762}$ & He~{\sc i}       & V2         & 2s 3S      & 3p 3P*     &          3 &        9    \\
        * &   3889.05 &  *     &             & *     &         & H~{\sc i}        & H8         & 2p+ 2P*    & 8d+ 2D     &          8 &        *    \\
  3919.70 &   3918.98 &    0.131& $\pm$   0.048&   0.175& $\pm$   0.050 & C~{\sc ii}       & V4         & 3p 2P*     & 4s 2S      &          2 &        2    \\
  3921.41 &   3920.69 &    0.303& $\pm$   0.045&   0.284& $\pm$   0.047 & C~{\sc ii}       & V4         & 3p 2P*     & 4s 2S      &          4 &        2    \\
  3927.26 &   3926.54 &    0.236& $\pm$   0.038&   0.253& $\pm$   0.039 & He~{\sc i}       & V58        & 2p 1P*     & 8d 1D      &          3 &        5    \\
  3965.38 &   3964.73 &    1.389& $\pm$   0.068&   1.390& $^{  +0.075}_{  -0.079}$ & He~{\sc i}       & V5         & 2s 1S      & 4p 1P*     &          1 &        3    \\
  3968.11 &   3967.46 &    2.465& $\pm$   0.247&   2.302& $^{  +0.257}_{  -0.260}$ & [Ne~{\sc iii}]   & F1         & 2p4 3P     & 2p4 1D     &          3 &        5    \\
  3970.73 &   3970.07 &   16.606& $\pm$   0.411&  17.675& $^{  +0.558}_{  -0.577}$ & H~{\sc i}        & H7         & 2p+ 2P*    & 7d+ 2D     &          8 &       98    \\
  3995.65 &   3994.99 &    0.074& $\pm$   0.026&   0.081& $\pm$   0.027 & N~{\sc ii}       & V12        & 3s 1P*     & 3p 1D      &          3 &        5    \\
  4009.92 &   4009.26 &    0.273& $\pm$   0.027&   0.284& $^{  +0.027}_{  -0.030}$ & He~{\sc i}       & V55        & 2p 1P*     & 7d 1D      &          3 &        5    \\
  4026.74 &   4026.08 &    3.395& $\pm$   0.088&   3.482& $^{  +0.059}_{  -0.087}$ & N~{\sc ii}       & V39b       & 3d 3F*     & 4f 2[5]    &          7 &        9    \\
        * &   4026.21 &  *     &             & *     &         & He~{\sc i}       & V18        & 2p 3P*     & 5d 3D      &          9 &       15    \\
  4041.98 &   4041.31 &    0.208& $\pm$   0.025&   0.169& $\pm$   0.025 & N~{\sc ii}       & V39b       & 3d 3F*     & 4f 2[5]    &          9 &       11    \\
  4044.20 &   4043.53 &    0.126& $\pm$   0.028&   0.092& $\pm$   0.029 & N~{\sc ii}       & V39a       & 3d 3F*     & 4f 2[4]    &          7 &        9    \\
  4085.78 &   4085.11 &    0.076& $\pm$   0.036&   0.111& $\pm$   0.037 & O~{\sc ii}       & V10        & 3p 4D*     & 3d 4F      &          6 &        6    \\
  4089.96 &   4089.29 &    0.121& $\pm$   0.026&   0.118& $\pm$   0.027 & O~{\sc ii}       & V48a       & 3d 4F      & 4f G5*     &         10 &       12    \\
  4097.93 &   4097.25 &    0.541& $\pm$   0.107&   0.554& $^{  +0.009}_{  -0.013}$ & O~{\sc ii}       & V48b       & 3d 4F      & 4f G4*     &          8 &       10    \\
        * &   4097.26 &  *     &             & *     &         & O~{\sc ii}       & V48b       & 3d 4F      & 4f G4*     &          8 &       10    \\
        * &   4097.33 &  *     &             & *     &         & N~{\sc iii}      & V1         & 3s 2S      & 3p 2P*     &          2 &        4    \\
  4102.39 &   4101.74 &   26.718& $\pm$   0.690&  27.326& $^{  +0.850}_{  -0.877}$ & H~{\sc i}        & H6         & 2p+ 2P*    & 6d+ 2D     &          8 &       72    \\
  4111.43 &   4110.78 &    0.134& $\pm$   0.045&   0.145& $\pm$   0.046 & O~{\sc ii}       & V20        & 3p  4P*    & 3d  4D     &          4 &        2    \\
  4119.87 &   4119.22 &    0.101& $\pm$   0.023&   0.128& $\pm$   0.024 & O~{\sc ii}       & V20        & 3p 4P*     & 3d 4D      &          6 &        8    \\
  4120.93 &   4120.28 &    0.249& $\pm$   0.023&   0.255& $^{  +0.004}_{  -0.006}$ & O~{\sc ii}       & V20        & 3p 4P*     & 3d 4D      &          6 &        6    \\
        * &   4120.54 &  *     &             & *     &         & O~{\sc ii}       & V20        & 3p 4P*     & 3d 4D      &          6 &        4    \\
        * &   4120.84 &  *     &             & *     &         & He~{\sc i}       & V16        & 2p 3P*     & 5s 3S      &          9 &        3    \\
  4122.11 &   4121.46 &    0.094& $\pm$   0.025&   0.117& $\pm$   0.026 & O~{\sc ii}       & V19        & 3p 4P*     & 3d 4P      &          2 &        2    \\
  4129.97 &   4129.32 &    0.052& $\pm$   0.015&   0.055& $\pm$   0.015 & O~{\sc ii}       & V19        & 3p 4P*     & 3d 4P      &          4 &        2    \\
  \hline
    \end{tabular}
 \end{table*}

 \begin{table*}
    \centering
    \caption{Table \ref{tab:lines1} continued.}
   \label{tab:lines2}
    \begin{tabular}{lrlrllllllll}
\hline
$ \lambda $ & Ion & $F \left( \lambda \right) $ && $I \left( \lambda \right) $ && Ion & Multiplet & Lower term & Upper term & g$_1$ & g$_2$ \\
\hline
  4133.45 &   4132.80 &    0.217& $\pm$   0.025&   0.212& $\pm$   0.026 & O~{\sc ii}       & V19        & 3p 4P*     & 3d 4P      &          2 &        4    \\
  4144.41 &   4143.76 &    0.493& $\pm$   0.022&   0.496& $\pm$   0.024 & He~{\sc i}       & V53        & 2p 1P*     & 6d 1D      &          3 &        5    \\
  4153.96 &   4153.30 &    0.276& $\pm$   0.021&   0.334& $^{  +0.021}_{  -0.023}$ & O~{\sc ii}       & V19        & 3p 4P*     & 3d 4P      &          4 &        6    \\
  4157.19 &   4156.53 &    0.078& $\pm$   0.023&   0.076& $\pm$   0.023 & O~{\sc ii}       & V19        & 3p 4P*     & 3d 4P      &          6 &        4    \\
  4169.63 &   4168.97 &    0.160& $\pm$   0.026&   0.163& $^{  +0.002}_{  -0.003}$ & He~{\sc i}       & V52        & 2p 1P*     & 6s 1S      &          3 &        1    \\
        * &   4169.22 &  *     &             & *     &         & O~{\sc ii}       & V19        & 3p 4P*     & 3d 4P      &          6 &        6    \\
  4190.45 &   4189.79 &    0.049& $\pm$   0.021&   0.065& $\pm$   0.021 & O~{\sc ii}       & V36        & 3p' 2F*    & 3d' 2G     &          8 &       10    \\
  4237.63 &   4236.91 &    0.126& $\pm$   0.020&   0.128& $\pm$   0.002 & N~{\sc ii}       & V48a       & 3d 3D*     & 4f 1[3]    &          3 &        5    \\
        * &   4237.05 &  *     &             & *     &         & N~{\sc ii}       & V48b       & 3d 3D*     & 4f 1[4]    &          5 &        7    \\
  4241.96 &   4241.24 &    0.186& $\pm$   0.026&   0.189& $^{  +0.002}_{  -0.004}$ & N~{\sc ii}       & V48a       & 3d 3D*     & 4f 1[3]    &          5 &        5    \\
        * &   4241.78 &  *     &             & *     &         & N~{\sc ii}       & V48b       & 3d 3D*     & 4f 1[4]    &          7 &        9    \\
  4267.87 &   4267.15 &    2.353& $\pm$   0.054&   2.493& $\pm$   0.066 & C~{\sc ii}       & V6         & 3d 2D      & 4f 2F*     &         10 &       14    \\
  4276.27 &   4275.55 &    0.179& $\pm$   0.019&   0.183& $^{  +0.002}_{  -0.003}$ & O~{\sc ii}       & V67a       & 3d 4D      & 4f F4*     &          8 &       10    \\
        * &   4275.99 &  *     &             & *     &         & O~{\sc ii}       & V67b       & 3d 4D      & 4f F3*     &          4 &        6    \\
        * &   4276.28 &  *     &             & *     &         & O~{\sc ii}       & V67b       & 3d 4D      & 4f F3*     &          6 &        6    \\
        * &   4276.75 &  *     &             & *     &         & O~{\sc ii}       & V67b       & 3d 4D      & 4f F3*     &          6 &        8    \\
  4295.50 &   4294.78 &    0.050& $\pm$   0.016&   0.051& $^{  +0.001}_{  -0.001}$ & O~{\sc ii}       & V53b       & 3d 4P      & 4f D2*     &          4 &        6    \\
        * &   4294.92 &  *     &             & *     &         & O~{\sc ii}       & V53b       & 3d 4P      & 4f D2*     &          4 &        4    \\
  4304.34 &   4303.61 &    0.276& $\pm$   0.018&   0.281& $^{  +0.003}_{  -0.005}$ & O~{\sc ii}       & V65a       & 3d 4D      & 4f G5*     &          8 &       10    \\
        * &   4303.82 &  *     &             & *     &         & O~{\sc ii}       & V53a       & 3d 4P      & 4f D3*     &          6 &        8    \\
  4317.87 &   4317.14 &    0.213& $\pm$   0.021&   0.216& $^{  +0.002}_{  -0.004}$ & O~{\sc ii}       & V2         & 3s 4P      & 3p 4P*     &          2 &        4    \\
        * &   4317.70 &  *     &             & *     &         & O~{\sc ii}       & V53a       & 3d 4P      & 4f D3*     &          4 &        6    \\
  4320.36 &   4319.63 &    0.065& $\pm$   0.016&   0.071& $\pm$   0.017 & O~{\sc ii}       & V2         & 3s 4P      & 3p 4P*     &          4 &        6    \\
  4341.15 &   4340.47 &   45.123& $\pm$   1.079&  46.500& $\pm$   1.200 & H~{\sc i}        & H5         & 2p+ 2P*    & 5d+ 2D     &          8 &       50    \\
  4350.12 &   4349.43 &    0.357& $\pm$   0.052&   0.310& $\pm$   0.053 & O~{\sc ii}       & V2         & 3s 4P      & 3p 4P*     &          6 &        6    \\
  4359.50 &   4358.81 &    0.057& $\pm$   0.015&   0.053& $\pm$   0.015 & [Fe~{\sc ii}]    & F7         & 3d6 3D     & 3d6 3P1    &          2 &        4    \\
  4363.90 &   4363.21 &    0.151& $\pm$   0.015&   0.175& $^{  +0.015}_{  -0.017}$ & [O~{\sc iii}]    & F2         & 2p2 1D     & 2p2 1S     &          5 &        1    \\
  4367.58 &   4366.89 &    0.196& $\pm$   0.018&   0.210& $^{  +0.018}_{  -0.019}$ & N~{\sc iii}      & V2         & 3s 4P      & 3p 4P*     &          6 &        4    \\
  4388.62 &   4387.93 &    0.730& $\pm$   0.028&   0.763& $\pm$   0.030 & He~{\sc i}       & V51        & 2p 1P*     & 5d 1D      &          3 &        5    \\
  4392.68 &   4391.99 &    0.076& $\pm$   0.021&   0.077& $^{  +0.001}_{  -0.001}$ & Ne~{\sc ii}      & V55e       & 3d 4F      & 4f 2[5]*   &         10 &       10    \\
        * &   4392.00 &  *     &             & *     &         & Ne~{\sc ii}      & V55e       & 3d 4F      & 4f 2[5]*   &         10 &       10    \\
  4417.67 &   4416.97 &    0.084& $\pm$   0.015&   0.104& $\pm$   0.015 & O~{\sc ii}       & V5         & 3s 2P      & 3p 2D*     &          2 &        4    \\
  4429.22 &   4428.52 &    0.053& $\pm$   0.016&   0.053& $^{  +0.001}_{  -0.001}$ & Ne~{\sc ii}      & V61b       & 3d 2D      & 4f 2[3]*   &          6 &        8    \\
        * &   4428.64 &  *     &             & *     &         & Ne~{\sc ii}      & V60c       & 3d 2F      & 4f 1[3]*   &          6 &        8    \\
  4431.64 &   4430.94 &    0.037& $\pm$   0.012&   0.044& $\pm$   0.012 & Ne~{\sc ii}      & V61a       & 3d 2D      & 4f 2[4]*   &          6 &        8    \\
  4433.44 &   4432.74 &    0.041& $\pm$   0.007&   0.042& $^{  +0.000}_{  -0.001}$ & N~{\sc ii}       & V55b       & 3d 3P*     & 4f 2[3]    &          5 &        7    \\
        * &   4432.75 &  *     &             & *     &         & N~{\sc ii}       & V55b       & 3d 3P*     & 4f 2[3]    &          5 &        7    \\
  4472.20 &   4471.50 &    6.121& $\pm$   0.144&   6.185& $\pm$   0.157 & He~{\sc i}       & V14        & 2p 3P*     & 4d 3D      &          9 &       15    \\
  4491.90 &   4491.07 &    0.118& $\pm$   0.013&   0.119& $\pm$   0.001 & C~{\sc ii}       &            & 4f 2F*     & 9g 2G      &         14 &       18    \\
        * &   4491.23 &  *     &             & *     &         & O~{\sc ii}       & V86a       & 3d 2P      & 4f D3*     &          4 &        6    \\
  4531.25 &   4530.41 &    0.093& $\pm$   0.013&   0.094& $^{  +0.001}_{  -0.001}$ & N~{\sc ii}       & V58b       & 3d 1F*     & 4f 2[5]    &          7 &        9    \\
        * &   4530.86 &  *     &             & *     &         & N~{\sc iii}      & V3         & 3s' 4P*    & 3p' 4D     &          4 &        2    \\
  4553.37 &   4552.53 &    0.069& $\pm$   0.017&   0.051& $\pm$   0.017 & N~{\sc ii}       & V58a       & 3d 1F*     & 4f 2[4]    &          7 &        9    \\
  4563.45 &   4562.60 &    0.039& $\pm$   0.015&   0.046& $\pm$   0.015 & Mg~{\sc i}]      &            & 3s2 1S     & 3s3p 3P*   &          1 &        5    \\
  4596.81 &   4595.96 &    0.044& $\pm$   0.010&   0.045& $^{  +0.000}_{  -0.000}$ & O~{\sc ii}       & V15        & 3s' 2D     & 3p' 2F*    &          6 &        6    \\
        * &   4596.18 &  *     &             & *     &         & O~{\sc ii}       & V15        & 3s' 2D     & 3p' 2F*    &          4 &        6    \\
  \hline
    \end{tabular}
 \end{table*}

 \begin{table*}
    \centering
    \caption{Table \ref{tab:lines1} continued.}
\label{tab:lines3}
    \begin{tabular}{lrlrllllllll}
\hline
$ \lambda $ & Ion & $F \left( \lambda \right) $ && $I \left( \lambda \right) $ && Ion & Multiplet & Lower term & Upper term & g$_1$ & g$_2$ \\
\hline
  4602.33 &   4601.48 &    0.128& $\pm$   0.016&   0.131& $\pm$   0.016 & N~{\sc ii}       & V5         & 3s 3P*     & 3p 3P      &          3 &        5    \\
  4607.88 &   4607.03 &    0.074& $\pm$   0.017&   0.074& $^{  +0.000}_{  -0.001}$ & [Fe~{\sc iii}]   & F3         & 3d6 5D     & 3d6 3F2    &          9 &        7    \\
        * &   4607.16 &  *     &             & *     &         & N~{\sc ii}       & V5         & 3s 3P*     & 3p 3P      &          1 &        3    \\
  4610.29 &   4609.44 &    0.081& $\pm$   0.014&   0.084& $\pm$   0.014 & O~{\sc ii}       & V92a       & 3d 2D      & 4f F4*     &          6 &        8    \\
  4613.86 &   4613.14 &    0.074& $\pm$   0.011&   0.074& $^{  +0.000}_{  -0.001}$ & O~{\sc ii}       & V92b       & 3d 2D      & 4f F3*     &          6 &        6    \\
        * &   4613.68 &  *     &             & *     &         & O~{\sc ii}       & V92b       & 3d 2D      & 4f F3*     &          6 &        8    \\
        * &   4613.87 &  *     &             & *     &         & N~{\sc ii}       & V5         & 3s 3P*     & 3p 3P      &          3 &        3    \\
  4621.98 &   4621.25 &    0.102& $\pm$   0.013&   0.103& $\pm$   0.001 & O~{\sc ii}       & V92        & 3d 2D      & 4f 2[2]*   &          6 &        6    \\
        * &   4621.39 &  *     &             & *     &         & N~{\sc ii}       & V5         & 3s 3P*     & 3p 3P      &          3 &        1    \\
  4631.27 &   4630.54 &    0.234& $\pm$   0.019&   0.259& $\pm$   0.019 & N~{\sc ii}       & V5         & 3s 3P*     & 3p 3P      &          5 &        5    \\
  4639.59 &   4638.86 &    0.305& $\pm$   0.016&   0.297& $\pm$   0.016 & O~{\sc ii}       & V1         & 3s 4P      & 3p 4D*     &          2 &        4    \\
  4642.54 &   4641.81 &    0.501& $\pm$   0.015&   0.504& $^{  +0.002}_{  -0.003}$ & O~{\sc ii}       & V1         & 3s 4P      & 3p 4D*     &          4 &        6    \\
        * &   4641.84 &  *     &             & *     &         & N~{\sc iii}      & V2         & 3p 2P*     & 3d 2D      &          4 &        4    \\
  4643.81 &   4643.08 &    0.066& $\pm$   0.014&   0.059& $\pm$   0.014 & N~{\sc ii}       & V5         & 3s  3P*    &  3p  3P    &          5 &        3    \\
  4649.86 &   4649.13 &    0.360& $\pm$   0.026&   0.388& $\pm$   0.026 & O~{\sc ii}       & V1         & 3s 4P      & 3p 4D*     &          6 &        8    \\
  4650.98 &   4650.25 &    0.283& $\pm$   0.026&   0.285& $^{  +0.001}_{  -0.002}$ & C~{\sc iii}      & V1         & 3s 3S      & 3p 3P*     &          3 &        3    \\
        * &   4650.84 &  *     &             & *     &         & O~{\sc ii}       & V1         & 3s 4P      & 3p 4D*     &          2 &        2    \\
  4652.20 &   4651.47 &    0.067& $\pm$   0.023&   0.105& $\pm$   0.023 & C~{\sc iii}      & V1         & 3s 3S      & 3p 3P*     &          3 &        1    \\
  4662.36 &   4661.63 &    0.272& $\pm$   0.018&   0.264& $\pm$   0.018 & O~{\sc ii}       & V1         & 3s 4P      & 3p 4D*     &          4 &        4    \\
  4674.46 &   4673.73 &    0.084& $\pm$   0.013&   0.083& $\pm$   0.013 & O~{\sc ii}       & V1         & 3s 4P      & 3p 4D*     &          4 &        2    \\
  4676.98 &   4676.24 &    0.131& $\pm$   0.015&   0.132& $^{  +0.015}_{  -0.016}$ & O~{\sc ii}       & V1         & 3s 4P      & 3p 4D*     &          6 &        6    \\
  4697.09 &   4696.35 &    0.077& $\pm$   0.015&   0.062& $\pm$   0.015 & O~{\sc ii}       & V1         & 3s 4P      & 3p 4D*     &          6 &        4    \\
  4713.91 &   4713.17 &    0.494& $\pm$   0.025&   0.457& $\pm$   0.025 & He~{\sc i}       & V12        & 2p 3P*     & 4s 3S      &          9 &        3    \\
  4802.65 &   4802.23 &    0.088& $\pm$   0.018&   0.088& $\pm$   0.000 & C~{\sc ii}       &            & 4f 2F*     & 8g 2G      &         14 &       18    \\
        * &   4803.29 &  *     &             & *     &         & N~{\sc ii}       & V20        & 3p 3D      & 3d 3D*     &          7 &        7    \\
  4861.76 &   4861.33 &  102.381& $\pm$   2.948& 100.000& $\pm$   3.000 & H~{\sc i}        & H4         & 2p+ 2P*    & 4d+ 2D     &          8 &       32    \\
  4891.29 &   4890.86 &    0.186& $\pm$   0.043&   0.136& $^{  +0.042}_{  -0.043}$ & O~{\sc ii}       & V28        & 3p 4S*     & 3d 4P      &          4 &        2    \\
  4922.37 &   4921.93 &    1.753& $\pm$   0.061&   1.759& $\pm$   0.060 & He~{\sc i}       & V48        & 2p 1P*     & 4d 1D      &          3 &        5    \\
  4959.35 &   4958.91 &   31.886& $\pm$   1.046&  32.300& $\pm$   1.000 & [O~{\sc iii}]    & F1         & 2p2 3P     & 2p2 1D     &          3 &        5    \\
  5007.28 &   5006.84 &   98.187& $\pm$   2.704&  97.900& $\pm$   2.700 & [O~{\sc iii}]    & F1         & 2p2 3P     & 2p2 1D     &          5 &        5    \\
  5197.97 &   5197.90 &    0.233& $\pm$   0.014&   0.227& $\pm$   0.014                                                                                      \\
  5200.33 &   5200.26 &    0.456& $\pm$   0.015&   0.444& $\pm$   0.015 & [N~{\sc i}]      & F1         & 2p3 4S*    & 2p3 2D*    &          4 &        6    \\
  5342.46 &   5342.38 &    0.133& $\pm$   0.018&   0.126& $\pm$   0.018 & C~{\sc ii}       &            & 4f 2F*     & 7g 2G      &         14 &       18    \\
  5453.91 &   5453.83 &    0.047& $\pm$   0.019&   0.075& $\pm$   0.018 & S~{\sc ii}       & V6         & 4s 4P      & 4p 4D*     &          6 &        8    \\
  5518.13 &   5517.66 &    0.222& $\pm$   0.017&   0.181& $^{  +0.016}_{  -0.018}$ & [Cl~{\sc iii}]   & F1         & 2p3 4S*    & 2p3 2D*    &          4 &        6    \\
  5538.07 &   5537.60 &    0.152& $\pm$   0.016&   0.159& $^{  +0.015}_{  -0.017}$ & [Cl~{\sc iii}]   & F1         & 2p3 4S*    & 2p3 2D*    &          4 &        4    \\
  5577.81 &   5577.34 &    0.067& $\pm$   0.014&   0.047& $\pm$   0.014 & [O~{\sc i}]      & F3         & 2p4 1D     & 2p4 1S     &          5 &        1    \\
  5667.04 &   5666.63 &    0.230& $\pm$   0.016&   0.214& $^{  +0.016}_{  -0.017}$ & N~{\sc ii}       & V3         & 3s 3P*     & 3p 3D      &          3 &        5    \\
  5676.44 &   5676.02 &    0.092& $\pm$   0.021&   0.108& $\pm$   0.021 & N~{\sc ii}       & V3         & 3s 3P*     & 3p 3D      &          1 &        3    \\
  5679.98 &   5679.56 &    0.324& $\pm$   0.018&   0.311& $\pm$   0.018 & N~{\sc ii}       & V3         & 3s 3P*     & 3p 3D      &          5 &        7    \\
  5686.63 &   5686.21 &    0.068& $\pm$   0.017&   0.075& $\pm$   0.017 & N~{\sc ii}       & V3         & 3s 3P*     & 3p 3D      &          3 &        3    \\
  5696.34 &   5695.92 &    0.033& $\pm$   0.009&   0.030& $\pm$   0.009 & C~{\sc iii}      & V2         & 3p 1P*     & 3d 1D      &          3 &        5    \\
  5711.19 &   5710.77 &    0.062& $\pm$   0.012&   0.076& $\pm$   0.011 & N~{\sc ii}       & V3         & 3s 3P*     & 3p 3D      &          5 &        5    \\
  5755.02 &   5754.60 &    0.586& $\pm$   0.024&   0.593& $\pm$   0.026 & [N~{\sc ii}]     & F3         & 2p2 1D     & 2p2 1S     &          5 &        1    \\
  5876.09 &   5875.66 &   21.589& $\pm$   0.855&  19.612& $\pm$   0.930 & He~{\sc i}       & V11        & 2p 3P*     & 3d 3D      &          9 &       15    \\
  5928.24 &   5927.81 &    0.063& $\pm$   0.010&   0.058& $\pm$   0.010 & N~{\sc ii}       & V28        & 3p 3P      & 3d 3D*     &          1 &        3    \\
  \hline
    \end{tabular}
 \end{table*}

 \begin{table*}
\label{tab:lines4}
    \centering
    \caption{Table \ref{tab:lines1} continued.}
    \begin{tabular}{lrlrllllllll}
\hline
$ \lambda $ & Ion & $F \left( \lambda \right) $ && $I \left( \lambda \right) $ && Ion & Multiplet & Lower term & Upper term & g$_1$ & g$_2$ \\
\hline
  5932.21 &   5931.78 &    0.110& $\pm$   0.011&   0.096& $^{  +0.010}_{  -0.011}$ & N~{\sc ii}       & V28        & 3p 3P      & 3d 3D*     &          3 &        5    \\
  5942.08 &   5941.65 &    0.114& $\pm$   0.012&   0.140& $^{  +0.012}_{  -0.013}$ & N~{\sc ii}       & V28        & 3p 3P      & 3d 3D*     &          5 &        7    \\
  6151.91 &   6151.43 &    0.068& $\pm$   0.020&   0.072& $\pm$   0.020 & C~{\sc ii}       &  V16.04    & 4d 2D      & 6f 2F*     &         10 &       14    \\
  6300.83 &   6300.34 &    0.977& $\pm$   0.040&   0.906& $\pm$   0.046 & [O~{\sc i}]      & F1         & 2p4 3P     & 2p4 1D     &          5 &        5    \\
  6311.29 &   6310.80 &    0.201& $\pm$   0.020&   0.194& $^{  +0.007}_{  -0.005}$                                                                                      \\
        * &   6312.10 &  *     &             & *     &         & [S~{\sc iii}]    & F3         & 2p2 1D     & 2p2 1S     &          5 &        1    \\
  6347.59 &   6347.10 &    0.065& $\pm$   0.007&   0.071& $^{  +0.006}_{  -0.007}$ & Si~{\sc ii}      & V2         & 4s 2S      & 4p 2P*     &          2 &        4    \\
  6364.27 &   6363.78 &    0.304& $\pm$   0.014&   0.289& $\pm$   0.016 & [O~{\sc i}]      & F1         & 2p4 3P     & 2p4 1D     &          3 &        5    \\
  6371.87 &   6371.38 &    0.068& $\pm$   0.014&   0.060& $\pm$   0.013 & S~{\sc iii}      & V2         & 4s 2S      & 4p 2P*     &          2 &        2    \\
  6463.07 &   6461.95 &    0.225& $\pm$   0.023&   0.194& $^{  +0.021}_{  -0.024}$ & C~{\sc ii}       &            & 4f 2F*     & 6g 2G      &         14 &       18    \\
  6549.24 &   6548.10 &   21.226& $\pm$   1.908&  23.500& $^{  +1.900}_{  -2.100}$ & [N~{\sc ii}]     & F1         & 2p2 3P     & 2p2 1D     &          3 &        5    \\
  6561.24 &   6560.10 &   70.074& $\pm$  23.415&  85.800& $^{ +22.400}_{ -22.500}$ & He~{\sc ii}      & 4.6        & 4f+ 2F*    & 6g+ 2G     &         32 &        *    \\
  6563.91 &   6562.77 &  299.373& $\pm$  13.157& 292.000& $^{  +7.000}_{  -6.000}$ & H~{\sc i}        & H3         & 2p+ 2P*    & 3d+ 2D     &          8 &       18    \\
  6584.64 &   6583.50 &   72.754& $\pm$   3.332&  67.700& $\pm$   3.900 & [N~{\sc ii}]     & F1         & 2p2 3P     & 2p2 1D     &          5 &        5    \\
  6679.32 &   6678.16 &    5.673& $\pm$   0.425&   5.417& $^{  +0.428}_{  -0.465}$ & He~{\sc i}       & V46        & 2p 1P*     & 3d 1D      &          3 &        5    \\
  6717.61 &   6716.44 &    5.551& $\pm$   0.401&   5.324& $^{  +0.410}_{  -0.444}$ & [S~{\sc ii}]     & F2         & 2p3 4S*    & 2p3 2D*    &          4 &        6    \\
  6731.99 &   6730.82 &    6.146& $\pm$   0.450&   5.830& $\pm$   0.473 & [S~{\sc ii}]     & F2         & 2p3 4S*    & 2p3 2D*    &          4 &        4    \\
  7065.70 &   7065.25 &    2.417& $\pm$   0.078&   2.277& $\pm$   0.115 & He~{\sc i}       & V10        & 2p 3P*     & 3s 3S      &          9 &        3    \\
  7136.25 &   7135.80 &    6.566& $\pm$   0.195&   5.695& $\pm$   0.292 & [Ar~{\sc iii}]   & F1         & 3p4 3P     & 3p4 1D     &          5 &        5    \\
  7161.05 &   7160.56 &    0.061& $\pm$   0.014&   0.055& $\pm$   0.014 & He~{\sc i}       &            & 3s 3S      & 10p 3P*    &          3 &        9    \\
  7231.81 &   7231.32 &    0.320& $\pm$   0.031&   0.339& $^{  +0.031}_{  -0.034}$ & C~{\sc ii}       & V3         & 3p 2P*     & 3d 2D      &          2 &        4    \\
  7236.68 &   7236.19 &    0.828& $\pm$   0.029&   0.785& $^{  +0.039}_{  -0.028}$ & C~{\sc ii}       & V3         & 3p 2P*     & 3d 2D      &          4 &        6    \\
        * &   7236.42 &  *     &             & *     &         & C~{\sc ii}       & V3         & 3p 2P*     & 3d 2D      &          4 &        6    \\
        * &   7237.17 &  *     &             & *     &         & C~{\sc ii}       & V3         & 3p 2P*     & 3d 2D      &          4 &        4    \\
        * &   7237.26 &  *     &             & *     &         & [Ar~{\sc iv}]    & F2         & 3p3 2D*    & 3p3 2P*    &          6 &        4    \\
  7281.84 &   7281.35 &    0.681& $\pm$   0.021&   0.641& $\pm$   0.033 & He~{\sc i}       & V45        & 2p 1P*     & 3s 1S      &          3 &        1    \\
  7298.54 &   7298.04 &    0.051& $\pm$   0.011&   0.042& $\pm$   0.010 & He~{\sc i}       &            & 3s 3S      & 9p 3P*     &          3 &        9    \\
  7319.42 &   7319.45 &    1.083& $\pm$   0.041&   1.026& $^{  +0.052}_{  -0.038}$ & [O~{\sc ii}]     & F2         & 2p3 2D*    & 2p3 2P*    &          6 &        2    \\
        * &   7319.99 &  *     &             & *     &         & [O~{\sc ii}]     & F2         & 2p3 2D*    & 2p3 2P*    &          6 &        4    \\
  7330.17 &   7330.20 &    0.410& $\pm$   0.042&   0.388& $^{  +0.020}_{  -0.014}$ & [O~{\sc ii}]     & F2         & 2p3 2D*    & 2p3 2P*    &          4 &        2    \\
        * &   7330.73 &  *     &             & *     &         & [O~{\sc ii}]     & F2         & 2p3 2D*    & 2p3 2P*    &          4 &        4    \\
 \hline
    \end{tabular}
 \end{table*}
\end{appendix}

\end{document}